\documentstyle[preprint,aps,eqsecnum,epsfig,rotating]{revtex}
\tighten
\newcommand{\teff}{T_{\hbox{\scriptsize eff}}}
\newcommand{\gr}[1]{{\bf #1}}
\newcommand{\ita}[1]{\hbox{\sl #1}}
\newcommand{\ch}{\,\hbox{ch}}
\newcommand{\sh}{\,\hbox{sh}}

\newcommand{\be}{\begin{equation}}
\newcommand{\ee}{\end{equation}}
\newcommand{\ve}[1]{\gr{#1}_{\gr{x}_0}}
\begin{document}
\title{\bf RENORMALIZATION GROUP FLOW AND FRAGMENTATION IN THE
SELF-GRAVITATING THERMAL GAS}   
\author{{\bf  B. Semelin$^{(b)}$, H. J. de Vega$^{(a)}$,
  N. S\'anchez$^{(b)}$ and F. Combes$^{(b)}$} \bigskip}
\address
{ (a)  Laboratoire de Physique Th\'eorique et Hautes Energies,
Universit\'e Paris VI, Tour 16, 1er \'etage, 4, Place Jussieu
75252 Paris, Cedex 05, FRANCE. Laboratoire Associ\'e au CNRS UMR 7589.\\
(b) Observatoire de Paris,  Demirm, 61, Avenue de l'Observatoire,
75014 Paris,  FRANCE.
Laboratoire Associ\'e au CNRS UA 336, Observatoire de Paris et
\'Ecole Normale Sup\'erieure.  \\ }
\maketitle
\begin{abstract}
The self-gravitating thermal gas (non-relativistic particles of mass $m$  at
temperature $T$) is exactly equivalent to a field theory with a single
scalar field $ \phi(\gr{x}) $  and 
exponential self-interaction. We build up  perturbation theory around a
 space dependent stationary point $ \phi_0(r) $ in a finite
size domain $ 
\delta \leq r \leq R , \; ( \delta<< R) $, which is relevant
for astrophysical applications (interstellar medium, galaxy
distributions). We compute the correlations of the gravitational
potential ($\phi$) and of the density 
 and find that they scale; the latter  scales as $ r^{-2} $. A
rich structure emerges in the two-point correlators from the $\phi$
fluctuations around $ \phi_0(r) $. The $n$-point correlators are
explicitly computed to the one-loop level. The relevant effective
coupling turns out to be $ \lambda = 4 \pi \; G \; m^2 / (T \; R)
$. The renormalization group equations (RGE) for the $n$-point
correlator are derived and the RG flow for the effective coupling $
\lambda(\tau), \; \tau = \ln(R/\delta)  $,  explicitly
obtained. A novel dependence on $ \tau $ emerges here. 
$ \lambda(\tau) $ vanishes  each time  $ \tau $
approaches   discrete values $ \tau=\tau_n = 2\pi n/\sqrt7-0 , \; n=0,\;
1, \; 2, \ldots  $. Such RG infrared stable behaviour [$
\lambda(\tau) $ decreasing with increasing $\tau$] is here connected
with low density self-similar fractal structures fitting one into another. For
scales  smaller than the points $ \tau_n $,  ultraviolet unstable
behaviour appears which we connect to Jeans' unstable behaviour,
growing density and fragmentation. Remarkably, we get a hierarchy of
scales and Jeans lengths following the geometric progression $ R_n =
R_0 \; e^{2\pi n /\sqrt7 } = R_0 \; [10.749087\ldots]^n $. A hierarchy of this
type is expected for non-spherical geometries, with a rate different
from $e^{2 n /\sqrt7 }$.
\end{abstract}

\section{Introduction and Results}

The statistical mechanics of the self-gravitating gas is a relevant
physical problem motivated by the physics of the cold 
interstellar medium (ISM) and the large scale galaxy
distributions\cite{nosprd,nosapj,nos98}. 

In the grand canonical ensemble, we showed\cite{nosprd} that the
self-gravitating gas at temperature $ T $ is exactly equivalent to a
field theory  
of a single scalar field $ \phi({\vec x}) $ with exponential
self-interaction through the {\bf local} euclidean action,
\be \label{accionI}
\displaystyle
S[\phi]\,=\, {1  \over \teff} \int\!\! d^3\!\gr{x} \left[ {1 \over 2} 
\left(\nabla \phi(\gr{x}) \right)^2 - \mu^2 e^{\phi(\gr{x})} \right].
\ee
where
$$
\mu^2 \,=\, \sqrt{ {2\over \pi} } \; \ita{z} \; G \; m^{7 \over 2} \;
\sqrt{T}\, \; , \qquad \teff \,=\, 4 \pi {G m^2 \over T} 
\,\, .
$$
Here $ m $ stands for the mass of the particles, $ G $ for
Newton's constant and $ z $ is the fugacity of the gas. The field $
\phi $ equals the gravitational potential up to a multiplicative constant. The
particle density is given by
$$
 \rho(\gr{r}) = -{1 \over \teff} \nabla^2 \phi(\gr{r})
$$

We analyzed this field theory perturbatively and non-perturbatively
through the renormalization group approach\cite{nosprd}. We showed
{\bf scaling}   
behaviour (critical) for a continuous range of the temperature and of 
the other physical parameters. We derived in this framework the scaling
relation  
$$ 
M(R) \sim R^{d_H} 
$$
 for the mass on a region of size $ R $,
and 
$$
 \Delta v \sim R^q 
$$ 
for the velocity dispersion where $ q = \frac12(d_H -1) $. For the
density-density correlations we found a power-law  behaviour for large
distances  
$$ 
\sim |{\vec r_1} -{\vec r_2}|^{2 d_H -6} \; .
$$
  The fractal dimension
$  d_H $ is related with the critical exponent $ \nu $ of the 
correlation length by 
$$  
d_H = 1/ \nu \; .  
$$
Mean field theory yields for the scaling
exponents $ \nu = 1/2  , \; d_H = 2 $ and $ q = 1/2 $. Such values
are compatible with the present ISM observational data: $  1.4    \leq
d_H    \leq   2     ,   \; 0.3  \leq     q  \leq 0.6 \;  $. 

In ref.\cite{nosapj} we developed a field theoretical approach to the
galaxy distribution. We considered a gas of self-gravitating masses 
in quasi-thermal equilibrium in the Friedman-Robertson-Walker background, 
We derived the galaxy correlations using renormalization group
methods. We found that the connected $N$-points density  correlator 
$ C({\vec r}_1,{\vec r}_2,\ldots,{\vec r}_N) $ scales as  
$$ 
r_1^{N(D-3)} \; , 
$$
when $ r_1 >> r_i, \; 2\leq i \leq N $. The theory is fully predictive
and there are no free parameters in it.

Our  study of the statistical mechanics of a self-gravitating system indicates 
that gravity provides a dynamical mechanism to produce   fractal
structure.

Besides the constant stationary point $ \phi_0 = - \infty $ studied in
ref.(\cite{nosprd}), the field theory defined by eq.(\ref{accionI})
possesses a rotationally invariant and dilatation invariant stationary point
\be \label{ficeroI}
\phi_0(\gr{r})\,=\, \ln { 2 \over \mu^2 \gr{r}^2}\; \; . 
\ee
Non-constant saddle points as $ \phi_0(\gr{r}) $ are clearly
neccesary in order to describe the physics of the self-gravitating
gas.

In the present paper we investigate the perturbation theory around the
stationary point given by eq.(\ref{ficeroI}) in the finite size domain
between a very large sphere of radius $ R $ and a small sphere of radius
$ \delta $. Such a finite domain $ \delta \leq r \leq R $ is dictated
by the physics of the problem. Both in the ISM as well as for the
galaxy distributions the relevant fractal domain is bounded; beyond $
R $ and below $ \delta $ other physics (besides purely gravitational)
do intervene. 

The stationary point  $ \phi_0(\gr{r}) $  may be centered at any
arbitrary point  $ \gr{x}_0 $ inside the domain considered. That is,
we must integrate over $ \gr{x}_0 $ treating it as a {\it collective
coordinate}\cite{cole}.

We explicitly compute the propagator of the field $ \phi $ (two points
correlator of the gravitational potential) in the background $
\phi_0(\gr{r}) $ to zeroth order [see 
eq.(\ref{propag})].  It presents a rich structure accounting for the
fluctuations around  $ \phi_0(\gr{r}) $.

The $n$-point correlation functions are explicitly computed 
at zero momentum to the
one-loop level. The perturbative series turns out to be an
expansion in the effective coupling
$$
\lambda \equiv {g^2 \over \mu R} = {4 \pi \; G \; m^2 \over T \; R} =
{\teff \over R}\; \; .
$$
Here, 
\be
g^2 = \mu\,  \teff = (8 \pi)^{3/4}\; \sqrt{z} \; \; {{G^{3/2}\;
m^{15/4}}\over T^{3/4}} 
\ee 
is the dimensionless coupling constant and $ \mu^{-1} = d_J $ is the
Jeans' length.  

The renormalization group equation (RGE) for $n$-point correlation functions
is derived and its coefficients computed to the order $ \lambda $. 
The RGE gives the variation of the correlation functions under a 
variation of scale. In the present case under variations  in $ R/\delta
$. We take here into account the explicit dependence on the size of
the system through the position of the boundaries. 

The new feature 
in the present model with respect to the standard cases \cite{itzdr}
is that the propagator has a non-trivial dependence on 
$ R / \delta $.  As a result, the coefficients in the RGE
have a novel non-trivial dependence on 
$$
 \tau \equiv \ln {R \over \delta} \; .
$$
We explicitly obtain the RG flow of the effective coupling $ \lambda =
\lambda(\tau) $. Integrating the RGE equation from a value $ \tau =
\tau_i $ where $ \lambda(\tau_i) = \lambda_i $ is small we find that $
\lambda(\tau) $ decreases until it vanishes at the first 
integer multiple of $ 2\pi/\sqrt7 $. That is at 
\be \label{jeraI} 
\tau_n = 2\pi n
/\sqrt7 , \; n=0,\; 1, \; 2, \ldots \; . 
\ee
Just after these points $
\tau =\tau_n $,  the one-loop approximation ceases to be valid since
the one-loop  $ \lambda(\tau) $ becomes negative. 

That is, we can start to run the renormalization group at $ \tau =
\tau_i $ with a small coupling $ \lambda_i \equiv
\lambda(\tau_i) $ and keep running until $ \tau = \tau_n $. At this
point $ \lambda(\tau_n) = 0 $. In this interval $ \tau_i,
\tau_n) $, the effective coupling  $ \lambda(\tau) $ decreases when
the spatial scale $ \tau $ increases as usually happens in scalar field 
theories, which is the case here ({\it infrared stable behaviour}). 

We depict in fig. 1 the running coupling constant  $ \lambda(\tau)
$ in the intervals $ (\tau_{n-1} , \tau_n) $ for $ n = 1, \; 2,  \; 3, \;
4, \; 6, \; 17 $ as illustrative cases.

From  eq.(\ref{jeraI}) we see that we have found a {\bf hierarchy} of 
 scales following the geometric progression
\begin{eqnarray}\label{jerarI}
R_0 &=& \delta \; ,\cr \cr
R_1 &=& R_0 \; e^{2\pi  /\sqrt7 }\; ,\cr \cr
\ldots && \ldots \cr \cr
R_n &=& R_0 \; e^{2\pi n /\sqrt7 } = R_0 \; [10.749087\ldots]^n\; .
\end{eqnarray}
We also expect hierarchies if this type for geometries different from the
spherical one. Of course, the rate $e^{2 /\sqrt7 }$ is expected to change
somewhat with the geometry.
In addition, $ \lambda(\tau) $ grows when $ \tau $ decreases starting from 
$ \tau = \tau_i $. This 
suggests that one enters here a strong coupling regime. That is, an
{\it ultraviolet unstable behaviour} that we connect with the
instability of structures and fragmentation.

As the effective coupling $\lambda(\tau)$ grows for decreasing $ \tau $ 
(ultraviolet unstable behaviour), the density fluctuations grow and the  
Jeans length $ d_J $
decreases, that is, smaller and smaller regions become unstable.
We thus get  {\bf fragmentation} of the original mass structure into
substructures.  

Remarkably, we get a {\bf hierarchy} of  scales  and a hierarchy of
Jeans lengths following the geometric progression 
$$
d_{J\; n} = d_{J\; 0} \; e^{2\pi n /\sqrt7 } \; n=0,\; 1, \; 2, \ldots \; . 
$$

This paper is organized as follows: in section II we summarize the
field theory approach to the gravitational gas, the properties of the
theory under scale transformations and the simplest relevant
stationary points. In section III we build up the perturbative series
around the space dependent stationary point $ \phi_0(\gr{r})
$. Section IV deals with the renormalization of the theory, the
derivation of the RGE, the RG flow of the effective coupling $
\lambda(\tau) $, physical results and their interpretation.

\section{The field theory approach to the gravitational gas}

As shown in ref.\cite{nosprd,nos98}, the statistical mechanics of a
gravitational gas 
can be described using a continuous field theory description. 

We consider a gas of nonrelativistic  point particles with mass $m$ 
interacting  only through their mutual gravitational attraction.
We ignore any relativistic or quantum
effect. Moreover, we assume this gas to be in thermal equilibrium at
temperature $ T $. We allow the number of particles
$N$ to vary through an exchange with the environment. In other words we 
 work in the grand canonical ensemble.

This system can then be described by its grand partition function, namely;

\be \label{parti}
{\cal Z}\,=\,\sum_{N=0}^{+\infty} {\ita{z}^N \over N!} \,\int\! \prod_{l=1}^{N}
{ d^3\!\gr{p}_l \; d^3\!\gr{q}_l \over (2 \pi)^3} \,\,\, e^{-\beta H_N},
\ee
where $\beta= {1 \over T}$ and

\be
H_N\,=\, \sum_{l=1}^{N}\, {\gr{p}_l^2 \over 2m} \,-\,\, G m^2\!\!\!
\sum_{1 \leq l < j \leq N}\, {1 \over | \gr{q}_l- \gr{q}_j |}\,\, ,
\ee
$G$ being Newton's constant and $\ita{z}$ the fugacity of the gas.

These N-body expressions can be exactly transformed into a continuum field
theoretical formula \cite{nosprd}. The number density field is given by

\be
\rho(\gr{r})\,=\, \sum_{j=1}^N \delta(\gr{r}-\gr{q}_j),
\ee
and the basic expression for the partition function is :
\begin{equation}
{\cal Z} \,=\, \int \!\! {\cal D} \phi \, \exp \left( - {1 \over {\teff}}
\int \!\! d^3\!\gr{x} \left[ {1 \over 2} ( \nabla \phi )^2 - \mu^2 e^{\phi} 
\right] \right) \; ,\label{Zintf}
\end{equation}
where

\be\label{param}
\mu^2 \,=\, \sqrt{ {2\over \pi} } \; \ita{z} \; G \; m^{7 \over 2} \;
\sqrt{T}\, \; , 
\qquad \teff \,=\, 4 \pi {G m^2 \over T} \qquad \mbox{and} \qquad 
\phi(\gr{x})=2m\sqrt{\pi G \beta} \; \psi(\gr{x})\,\, .
\ee

This expression of the partition function leads to a new insight on 
the self-gravitating gas. Eqs.(\ref{parti}) and eq.(\ref{Zintf}) are exactly
equivalent. Indeed, the system is exactly described by 
a single scalar field $\phi(\gr{x})$ with  {\bf local} euclidean action,

\be \label{accion}
\displaystyle
S[\phi]\,=\, {1  \over \teff} \int\!\! d^3\!\gr{x} \left[ {1 \over 2} 
\left(\nabla \phi(\gr{x}) \right)^2 - \mu^2 e^{\phi(\gr{x})} \right].
\ee
One characteristic feature of this action is that it is unbounded from below 
due to the sign in front of the exponential self-interaction term. 
This precisely reflects the presence of an attractive force like
 gravitation. It is then possible that some states  evolve
into gravitational collapse. To avoid the 
divergencies linked to gravitational collapse we  introduce a regularizing
cutoff at short distances. A simple short distance regularization of the 
Newtonian force for the two-body potential is\cite{nosprd}
\be\label{potcut}
v_a(\gr{r}) = -{{G m^2} \over r}\; [ 1 - \theta(a-r) ] \; ,
\ee
$ \theta(x)$ being the step function. The value of the lower cutoff $ a $
depends on the physical problem under consideration.

\bigskip

It is useful to study the mean field equation for the stationary points of 
the local action. It takes the form,

\be\label{eqclas}
\nabla^2 \phi_0(\gr{x})\,+\, \mu^2 e^{\phi_0(\gr{x})} \,=\,0.
\ee
 
We give below an equivalence table 
between the $N$-body description and the field theoretical formulation.

\bigskip

\begin{eqnarray}\label{tabla}   
\qquad \hbox{$N$ point particles} &\qquad&  \hbox{Continuous scalar
field} \nonumber 
\\ & &\nonumber \\ 
 U(\gr{r})= -\,\, G m
\sum_{1 \leq l < j \leq N}\, {1 \over | \gr{q}_l- \gr{q}_j |} \quad
&\longrightarrow& \quad -{T \over m} \phi(\gr{r})\nonumber  \\
&&\nonumber \\
 \rho(\gr{r})=  \displaystyle \sum_{j=1}^N \delta(\gr{r}-\gr{q}_j)\quad
&\longrightarrow& \quad -{1 \over \teff} \nabla^2 \phi(\gr{r})\nonumber  \\
&&\nonumber \\
 4\pi {G m^2 \over T}  \quad  &\longrightarrow &\quad\teff\nonumber  \\ 
\end{eqnarray}

\subsection{Behavior under scale transformation}

Let us briefly study the behaviour of the exponential field theory
eq.(\ref{accion}) under scale transformations. This is a preliminary 
step to the investigation the renormalization group in sec. IV. 
Scale transformations are defined by
\be
\gr{r} \stackrel{T_{\lambda}}{\longrightarrow} \gr{r}_{\lambda}\,=\, 
\lambda\gr{r}\,\, ,
\ee
where $\lambda$ is a real number. Sometimes we will consider $\lambda=1+
\epsilon$ close to 1. Under $T_{\lambda}$  the field $\phi$ transform as

\be \label{trafi}
\phi(\gr{r}) \stackrel{T_{\lambda}}{\longrightarrow} \phi_{\lambda}(\gr{r})=
\phi(\lambda\gr{r})+\ln \lambda^2.
\ee

\noindent 
And the classical field equation (\ref{eqclas}) remains.

Since $\phi$ is the gravitational potential the addition of the 
constant $ \ln \lambda^2 $ does not alter the physics.

Our field theory approach can be also generalized to a
$D$ dimensional space. The lagrangian density ${\cal L}$ retain the 
same structure with 
$D$ dimensional operators (see \cite{nosprd} for details), 

\be
{\cal L}[\phi]\,=\, {1 \over 2} \left( \nabla_D \phi \right)^2- \mu^2 e^{\phi},
\ee
where $\nabla_D$ is the $D$ dimensional gradient.

The scale transformations  $T_{\lambda}$ do not change and 
the action undergoes the following transformation under $T_{\lambda}$,

\be\label{trafoS}
S[\phi] \stackrel{T_{\lambda}}{\longrightarrow} S[\phi_{\lambda}] \,=\,
\lambda^{2-D}\, S[\phi].
\ee
The Noether current associated with this transformation is

\be
\gr{J}(\gr{r})\,=\, (\gr{r}.\nabla \phi +2)\nabla \phi(\gr{r})\,-\,\gr{r}
{\cal L}.
\ee
Its divergence in agreement  with eq.(\ref{trafoS}), is given by 

\be
\nabla {\gr{J}}\,=\, (2-D){\cal L}.
\ee

We did not expect a cancellation of this divergence since the action is
{\sl not} invariant under $T_{\lambda}$ except in two space dimensions.
One of the important features of this study is that, despite the presence 
of a characteristic length $\mu$ in the lagrangian density, the action {\bf 
scales} under a {\bf scale} transformation. 

\subsection{Stationary points}
 
We consider the self-gravitating gas in a large but finite spherical box
of radius $R$. In addition, we introduce a short distance cutoff at distance
$ r = \delta $. This cutoff $  \delta \sim a $ is the same as the
short-distance cutoff in the gravitational force [eq.(\ref{potcut})].

In order to extract information from the  functional integral 
eq.(\ref{Zintf}) we shall search for stationary points of the action
eq.(\ref{accion}). 

Let $\phi_0$ be a stationary point of the action. This means that $\phi_0$
is solution of the equation
 
\be \nabla^2 \phi(\gr{r})+ \mu^2 e^{\phi(\gr{r})}\,=\,0 \ee
It is interesting to notice that the dilated field $\phi_{\lambda}$ is also a 
stationary point. 

The most simple stationary point is \cite{nosprd}

\be \phi_0(\gr{r}) \,=\, -\infty, \ee
which describes  empty space. This is a translationally invariant stationary 
point. It has been studied in ref.\cite{nosprd}. 

Let us now study the  non-translationally invariant but 
rotationally invariant stationary  point 

\be \label{ficero}
\phi_0(\gr{r})\,=\, \ln { 2 \over \mu^2 \gr{r}^2}\; \; . 
\ee
This solution is also dilatation invariant under the transformation 
(\ref{trafi}). 

Less symmetrical stationary points certainly exist but their
calculation would require  numerical investigation  of the stationary
field equation.

\subsection{Hydrostatic interpretation of the stationary points}

We recall here that there is a straightforward correspondence between
our stationary point equation (\ref{eqclas}) and the equations for the 
self-gravitating  fluid in hydrostatic equilibrium.
In terms of the gravitational potential $U(\gr{x})$ [see eq.(\ref{tabla})],
eq.(\ref{eqclas})  takes the form

\begin{equation}\label{equih}
\nabla^2U(\gr{r}) = 4 \pi G \, z \,  m
\left({{mT}\over{2\pi}}\right)^{3/2} \,  e^{ - \frac{m}{T}\,U(\gr{r})} \; .
\end{equation}
This corresponds to the Poisson equation for a thermal matter distribution
fulfilling the hydrostatic equilibrium of an ideal gas.
The hydrostatic equilibrium condition is \cite{sas}
$ \nabla P(\gr{r}) = - m \, \rho(\gr{r}) \; \nabla U(\gr{r})\; , $
where $ P(\gr{r}) $ stands for the pressure,and the equation of state for the 
ideal gas $ P = T \rho \;$ yields for the particle density $
 \rho(\gr{r}) =  \rho_0 \; e^{ - \frac{m}{T}\,U(\gr{r})} \; $,
where $ \rho_0 $ is a constant. Inserting this relation into the
Poisson equation $ \nabla^2U(\gr{r}) = 4 \pi G\, m \, \rho(\gr{r}) $
yields eq.(\ref{equih}) with 
\begin{equation} \label{RO0}
  \rho_0 =  z \,\left({{mT}\over{2\pi}}
\right)^{3/2} \; . 
\end{equation}

The solution $\phi_0$ given by eq.(\ref{ficero}) is singular at $ r = 0$. 
However, such short distance  singularity 
is inessential for the long distance properties we are
investigating. There exist also spherically symmetric regular solutions 
known as the {\sl isothermal sphere} \cite{chan}. 

\section{Perturbation theory around a non constant stationary point} 

We will build our perturbation theory around the stationary point 
$ \phi_0 $ which is invariant under rotations and dilations. That is,
\be
\phi_0(\gr{x}-\gr{x}_0) \,=\, \ln \left( { 2 \over \mu^2
|\gr{x}-\gr{x}_0|^2 } \right) \;. 
\ee
This stationary point is centered in an arbitrary point $ \gr{x}_0 $.

\subsection{Construction of the perturbation series}

As usual in field theory,
we will introduce an external source $ {\sqrt{\mu} \over g} J(\gr{x})
$ in order  to define the correlation functions:

\be
S[\phi]= {1 \over \teff}\int \!\! d^3\!\gr{x} \left[{1 \over 2} (
\nabla_{\gr{x}} \phi )^2 -\mu^2 e^{\phi}\right] - \int  d^3\!x \;
{\sqrt{\mu} \over g}\,J(\gr{x})\; \phi(\gr{x}) \; .
\ee
$J({\gr{x}})$ can be interpreted as a test-mass density. Anyway, one
sets it equal to zero when computing physical quantities. 

Let us now perturb around the stationary point (\ref{ficero})

\be
\phi(\gr{x}) \,=\, \phi_0(\gr{x}-\gr{x}_0) + {g \over \sqrt{\mu} } \; 
\chi(\gr{x}-\gr{x}_0) \; ,
\ee
where 
\be\label{mutef}
g \equiv \sqrt{\mu \teff} 
\ee
 is the dimensionless coupling constant
and $\chi(\gr{x})$ is, of course, the fluctuation. 
The action can be then splitted into several terms,

\be
S[\phi]=S[\phi_0]+S_2[g\chi]+S_I[g\chi]-  \int\!  d^3\!x \;
J({\gr{x}})\; \chi({\gr{x}}) \;, 
\ee
where  $ S_2 $ stands  for the quadratic action and  $ S_I $ for 
higher order couplings,
\be
S_2[\chi]\,=\,  {1\over 2}\int\!\! d^3x\, \left[
(\nabla \chi(\gr{x}))^2 - \mu^2 e^{\phi_0(\gr{x})}  \; \chi^2(\gr{x}) \right]
\; ,
\ee
and
\be
S_I[\chi]\,=\, - \int\!\! d^3\!x \, \mu^2 e^{\phi_0(\gr{x})}
\sum_{n=3}^{\infty}{ g^{n-2} \; \chi^n({\gr{x}}) \over \sqrt{\mu}^{n-2}\; n! }\; .
\ee

We have a non-polynomial interaction action, hence we shall have 
vertices with any number of legs in the Feynman graphs.

We compute now the field propagator in the background of the solution
$\phi_0(\gr{x})$ as the inverse of the quadratic form in  $ S_2 $.

\subsection{Calculation of the propagator ${\cal G}$ in the 
$\phi_0$ background}

We read the inverse propagator  ${\cal G}^{-1}$ from the  quadratic
action $ S_2 $,

\be 
{\cal G}^{-1}(\gr{x}-\gr{x}_0,\gr{y}-\gr{x}_0)=  \delta(\gr{x}-\gr{y})
\left[ -\nabla_{\gr{x}-\gr{x}_0}^2
-\mu^2 e^{\phi_0(|\gr{x}-\gr{x}_0|)} \right] .
\ee
The propagator ${\cal G}$ is defined by the inversion relation:

\be
\int d^3z \,\,{\cal G}^{-1}(\gr{x}-\gr{x}_0,\gr{z}-\gr{x}_0)
{\cal G}(\gr{z}-\gr{x}_0,\gr{y}-\gr{x}_0)= \delta^3(\gr{x} -\gr{y}).
\ee
We can recast this definition as a partial differential equation,

\be 
\left[ -\nabla_{\gr{x}-\gr{x}_0}^2-\mu^2e^{\phi_0(|\gr{x}-\gr{x}_0|)} \right]
{\cal G}(\gr{x}-\gr{x}_0,\gr{y}-\gr{x}_0)=\delta^3 (\gr{x}-\gr{y}).
\ee
Using the explicit form of $\phi_0$ and the notation 
$\gr{r}=\gr{x}-\gr{x}_0 $;
 $\gr{r}^{\prime}=\gr{y}-\gr{x}_0$, we find

\be \label{laplaG}
\left[- \nabla^2 - {2 \over \hbox{\bf r}^2} \right] {\cal G}(\hbox{\bf r}
,\hbox{\bf r}^{\prime})=\delta(\hbox{\bf r}-\hbox{\bf r}^{\prime}).
\ee
It is convenient to expand the correlation function  in partial waves

\be
{\cal G}(\hbox{\bf r},\hbox{\bf r}^{\prime})=\sum_{l=0}^{\infty}
g_l(r,r^{\prime}) \sum_{m=-l}^{m=+l}
Y_{l,m}(\theta,\phi) \;  Y_{l,m}^\star (\theta^\prime,\phi^\prime)
\ee
where $ Y_{l,m}(\theta,\phi) $ are harmonic spherics and 
the function $ g_l(r,r^{\prime}) $ obey the ordinary differential 
equation

\bigskip
\noindent
\be
\displaystyle
\left[ - {d^2 \over dr^2}- {2 \over r} {d \over dr} + 
{ {l (l+1)-2} \over r^2} \right] g_l(r,r^{\prime}) = {1 \over r^2}\;
\delta(r-r^{\prime}) \; . 
\ee

\bigskip

The general solution of the homogeneous differential equation
\be \label{echom}
\left[ - {d^2 \over dr^2}- {2 \over r} {d \over dr} + 
{ {l (l+1)-2} \over r^2} \right] y_l(r) = 0
\ee
can be written as

$$
y_l(r) = {1 \over {\sqrt{r}}} \left( A_l \; r^{\alpha_l}
+ B_l \;  r^{-\alpha_l} \right)
$$
where $  A_l $ and $ B_l $ are arbitrary constants and
\be\label{alfa}
\alpha_l=\sqrt{\left(l+\frac12\right)^2-2}=\sqrt{l^2+l - \frac{7}{4}} \; .
\ee

As stated in the previous section we consider the gas in a finite volume
of radius $ R $ with a short distance cutoff $ \delta \sim a
$. Therefore, we need  solutions $ g_l(r,r^{\prime}) $ in the interval
$$
 \delta \leq r \leq R ,\; \delta \leq r^{\prime} \leq R \; .
$$

The radial propagator  $ g_l(r,r^{\prime}) $ can be then expressed in terms
of the solutions of the homogeneous equation (\ref{echom}) as follows:
\be
 g_l(r,r^{\prime}) = {1 \over { r^2 \; W[y_l^>,y_l^<] }} \;
y_l^<(r_<) \; y_l^>(r_>) \; .
\ee
where 
$$ 
r_< \equiv min(r,r^{\prime}) ,\;  r_> \equiv max(r,r^{\prime}) \; ,
$$
$ y_l^<(r) $ obeys the boundary conditions appropriate for $ r = \delta $ and
$  y_l^>(r) $ obeys the boundary conditions appropriate for $ r = R $.
$ W[y_l^>,y_l^<] $ stands for the wronskian. Notice that 
$ r^2\;  W[y_l^>,y_l^<] $ is $r$-independent.

The simplest choice of boundary  conditions is to set the fluctuations equal to
zero both at $  r= \delta $ and at $ r = R $.  That is,

\be
y_l^<(\delta) = 0 \quad , \quad  y_l^>(R) = 0 \; .
\ee

The radial part of the propagator is now fully determined, it takes the  form,

\be
\displaystyle
g_l(r,r^\prime)=-{1 \over \alpha_l \sh\left(\alpha_l \tau\right) }
{1 \over \sqrt{r r^\prime} }\sh\left[\alpha_l\ln \left( {r^< \over \delta}
\right) \right]
\sh\left[ \alpha_l\ln\left( { r^> \over R } \right)\right]\; ,
\ee
where 
$$
\tau\equiv \ln{R \over \delta} \; .
$$
Notice that for $ l = 0 , \; \alpha_0$ being [eq.(\ref{alfa})] an imaginary
number, the hyperbolic sinus turn into trigonometric sines. 
This is  an important point since  S-waves 
produce the leading contribution in loop calculations as we shall see in 
the subsequent sections.
\be
\displaystyle
g_0(r,r^\prime)=-{1 \over {\frac{\sqrt7}{2}
 \sin\left(\frac{\sqrt7}{2} \tau\right) }}
{1 \over \sqrt{r r^\prime} }
\sin\left[\frac{\sqrt7}{2}\ln \left( {r^< \over \delta}
\right) \right]
\sin\left[ \frac{\sqrt7}{2}\ln\left( { r^> \over R } \right)\right]\;,
\ee

We finally get for the full propagator:

\begin{eqnarray}\label{propag}
\displaystyle
{\cal G}(\hbox{\bf r},\hbox{\bf r}^\prime)&=&
{1 \over  {4\pi\sqrt{ r \, r'}}} \,\left\{ -{2 \over \sqrt7
 \sin\left(\frac{\sqrt7}{2} \tau\right) }
\sin\left[\frac{\sqrt7}{2}\ln \left( {r^< \over \delta}
\right) \right]
\sin\left[ \frac{\sqrt7}{2}\ln\left( { r^> \over R } \right)\right]
\right.
\cr \cr
& - & \left. \sum_{l\geq 1} {{2l+1} \over {\alpha_l \sh(\alpha_l \tau)}} \;
\sh\left[\alpha_l\ln\left({r^> \over R}\right)\right]
\sh\left[\alpha_l\ln\left({r^< \over \delta}\right)\right]
P_l(\cos\gamma) \right\} \; .
\end{eqnarray}
where we used the relation
$$
\sum_m Y_{l,m}(\theta,\phi)\; Y_{l,m}^\star (\theta^\prime,\phi^\prime)\,=\,
{{ 2l+1} \over 4\pi}\; P_l(\cos\gamma)\; .
$$
and $  P_l(\cos\gamma) $ stand for Legendre polynomials.

We can see that the propagator (\ref{propag2}) presents a rich  structure. 
It takes into account the behaviour of the fluctuations around the background
$\phi_0$.
Notice that the propagator  (\ref{propag2}) is 
symmetric under the  exchange of $\gr{r}$ and $\gr{r}^{\prime}$. 

As one sees from eq.(\ref{propag2}) for distances such as $ \delta \ll r \sim
r' \ll R $ the two-point function of the gravitational potential scales as 
$ \sim 1/\sqrt{r\,r'} $ at the tree level. 

Besides if  $ \delta \ll r \ll r' \ll R$ the propagator $ {\cal
G}(\hbox{\bf r},\hbox{\bf r}^\prime) $ is dominated by the S-wave. That is,
$$
{\cal G}(\hbox{\bf r},\hbox{\bf r}^\prime)
 {\buildrel{ r, \, r' >> \delta }\over =}-{1 \over  {2\pi\sqrt7\;\sqrt{ r \,
 r'}}}\,
 \sin\left(\frac{\sqrt7}{2} \tau\right) 
\sin\left[\frac{\sqrt7}{2}\ln \left( {r^< \over \delta}
\right) \right]
\sin\left[ \frac{\sqrt7}{2}\ln\left( { r^> \over R } \right)\right]
$$

\subsection{Perturbation theory, Generating functionals and
Feynman graphs}

We construct now the perturbative series around the stationary point
$ \phi_0 $. We use standard perturbation theory methods\cite{itzdr}. We
shall  give the main steps and the new features which appear
 for our model.

\subsubsection{Construction of the general Green functions}

We can write the generating  functional of disconnected correlations 
as follows
\be \label{zeta}
Z[J]\,=\, \int {d^3 \gr{x}_0 \over {\cal N} } \; Z_{\gr{x}_0}[J]
\ee

\be \label{zetaJ}
 Z_{\gr{x}_0}[J]\,=\,
\,e^{\int \!\! d^3\!\gr{x}\, \phi_0(\gr{x}-\gr{x}_0)
J(\gr{x}) {\sqrt{\mu} \over g } }
\int\!\! D \chi \; e^{-S_2[\chi]-S_I[\chi]+g \int  d^3\!\gr{x} \;
J(\gr{x})\; \chi(\gr{x}-\gr{x}_0)} \; 
\ee

[The name disconnected arises because topologically disconnected
Feynman graphs contibute to the perturbative series  of $ Z[J] $.] 

Here we used that $S[\phi_0]=0$ in three space dimensions. 
Notice first that the generating functional is now
a functional of $ J(.) $. For $J=0$ we recover the previous definition
eq.(\ref{parti}).

As we have already mentioned,  
$\phi_0(\gr{x}-\gr{x}_0)$ is a stationary point for any $\gr{x}_0$.
We have therefore to integrate over $ \gr{x}_0 $.  That is,
$ \gr{x}_0 $ must be treated  as a so called  {\sl collective
coordinate}\cite{cole}.
Such integration  over  $ \gr{x}_0 $ must
be computed exactly since all points $ \gr{x}_0 $ contribute with similar
weight due to the translational invariance within the large sphere of 
radius $ R $. In order to avoid double counting, the domain of
functional integration for the field $ \chi(\gr{x}) $ must be
orthogonal to the linear space of the zero modes.
That is, we have to impose the constraints \cite{cole}
\be\label{vinc}
\int \!\! d^3\!\gr{x} \,\,\chi(\gr{x}) \; \partial_{x_i} \phi_0(
\gr{x}-\gr{x}_0)=0 \quad \mbox{for}\; i = 1,2,3 \quad .
\ee
 ${\cal N}$ is a normalization factor coming from the Jacobian,
\be \label{jaco}
{\cal N}=\left[ {1 \over 6\pi} \int \!\! d^3 \! \hbox{\bf r} \left( \nabla
\phi_0 \right)^2 \right]^{3 \over 2}\,=\,\left[{ 8 \over 3} (R-\delta)
\right]^{3 \over 2}.
\ee

Since $ \phi_0(x) $ is rotationally invariant,
$\partial_{x_i}\phi_0(x), \; i=1,2,3 $ can be expressed as linear
combinations of the $ l=1 $ modes of the previous section. Namely, $
y_1(r)\,  Y_{1,\pm 1}(\theta,\phi) $ and $ y_1(r)\,
Y_{1,0}(\theta,\phi) $. The constraint (\ref{vinc}) implies that we
must exclude the $ l = 1 $ term in the partial wave series. We shall
denote from now on, 

\begin{eqnarray}\label{propag2}
\displaystyle
{\cal G}(\hbox{\bf r},\hbox{\bf r}^\prime)&=&
{1 \over  {4\pi\sqrt{ r \, r'}}} \,\left\{ -{2 \over \sqrt7
 \sin\left(\frac{\sqrt7}{2} \tau\right) }
\sin\left[\frac{\sqrt7}{2}\ln \left( {r^< \over \delta}
\right) \right]
\sin\left[ \frac{\sqrt7}{2}\ln\left( { r^> \over R } \right)\right]
\right.
\cr \cr
& - & \left. \sum_{l\geq 2} {{2l+1} \over {\alpha_l \sh(\alpha_l \tau)}} \;
\sh\left[\alpha_l\ln\left({r^> \over R}\right)\right]
\sh\left[\alpha_l\ln\left({r^< \over \delta}\right)\right]
P_l(\cos\gamma) \right\}\; .
\end{eqnarray}
 
It is convenient to split the exponent in eq.(\ref{zetaJ}) in two 
parts, separating  the interaction part of the action from the
gaussian part that can be easily integrated,

\be
\int \!D\chi\, e^{- \frac12 \int  d^3\!x\,\chi\,{\cal G}^{-1}\,\chi\,  +
 \int  d^3\!x \;
J(\gr{x})\; \chi(\gr{x}-\gr{x}_0)} \; =\,
\sqrt{\hbox{det}{\cal G}} \,\,e^{
{1 \over 2 } \int d^3\!x  d^3\!y \, J(\gr{x})\,
{\cal G}(\gr{x}-\gr{x}_0,\gr{y}-\gr{x}_0)  \, J(\gr{y})}\; .
\ee
Using this result and expanding exponentials,  one gets after calculation,

\begin{eqnarray}
&&
Z_{\gr{x}_0}[J]\,=\, \sqrt{\hbox{det}{\cal G}} 
\, e^{{\sqrt{\mu} \over g} \int  d^3\!x \; J \phi_0  }
\,\,\sum_{n=0}^{\infty} {1 \over n!}
\left( - S_I\left[{\delta \over \delta J(\gr{x})} \right] \right)^n
\nonumber \\ 
&& \nonumber \\
&& \qquad \qquad \qquad \qquad \qquad \qquad \qquad
\times \sum_{k=0}^{\infty} { 1 \over k!} \left( {1 \over 2}
\int \int d^3\!x\, d^3\!y \; J(\gr{x})\;  
{\cal G} (\gr{x}-\gr{x}_0,\gr{y}-\gr{x}_0)\; J(\gr{y}) \, \right)^k
\; . \nonumber \\
\end{eqnarray}

$ Z_{{\gr{x}}_0}[J]$ is 
thus expressed as a power series of  $J$. The coefficients of the 
$n$-th degree term is  the disconnected $n$ points correlation
function of the  field $ \phi $ (i.e the Green's function). 

That is,

\be
G_{\gr{x}_0}^{(n)}(\gr{x}_1,\ldots,\gr{x}_n)=
{1 \over Z_{\gr{x}_0}[0]}\,\,\,\,\left. { {\delta^n Z_{\gr{x}_0}[J]}
\over {\delta J(\gr{x}_1) ... \delta J(\gr{x}_n)} }\, \right|_{J=0}.
\ee
We will compute  $G^{(2)}$ and $G^{(1)}$ to the first
orders in $g$ in terms of  Feynman  diagrams.

\subsubsection{ The two points Green function and the Feynman rules}

Let us first compute the sum of  ``vacuum-vacuum'' diagrams (without
external sources $J$) $Z_{\gr{x}_0}[0]$.

\bigskip
\begin{eqnarray}
\displaystyle
Z_{\gr{x}_0}[0]\,&=&\, 1\, + \,{ g^2 \over 8\; \mu }\int \! d^3\!\gr{x} \,
\mu^2 \,  e^{\phi_0(\ve{x})} \left[{\cal G}(\ve{x},\ve{x}) \right]^2 \cr \cr
&-&\, {{g^4}\over 32\; \mu^2} \int \! d^3\!\gr{x} \,\mu^2\, e^{\phi_0(\ve{x})}
\left[ {\cal G}(\ve{x},\ve{x})\right]^3 \, + {\cal O}(g^6) 
\end{eqnarray}

\bigskip
\noindent
Here we  use the notation $ \ve{r}=\gr{r}-\gr{x}_0 $. Then, the  one and
two points correlation functions can be written as,
\be
G_{\gr{x}_0}^{(1)}(\gr{x})=\phi_0(\ve{x})
\, + \, {1 \over Z_{\gr{x}_0}[0] } \; g^{(1)}_{\chi}( \ve{x})\;,
\ee

\be
G_{\gr{x}_0}^{(2)}(\gr{x},\gr{y})=
\phi_0(\ve{x}) \phi_0(\ve{y})+
{ 1\over Z_{\gr{x}_0}[0]  } \left[ 
\phi_0(\ve{x})g^{(1)}_{\chi}(\ve{y}) 
+ \phi_0(\ve{y}) g^{(1)}_{\chi}(\ve{x}) + g^{(2)}_{\chi}(\ve{x},\ve{y}) 
\right],
\ee
where,
\begin{eqnarray}
&&
g^{(1)}_{\chi}(\ve{y})=-{g \over \sqrt{\mu}\;2} \int \! d^3\!x
\,\mu^2\; 
e^{\phi_0(\ve{x})}\;  {\cal G}(\ve{x},\ve{x})\;  {\cal G}(\ve{x},\ve{y})
\nonumber \\
&& \nonumber \\
&& \qquad \qquad \qquad \qquad \qquad \qquad
+ {g^3 \over 8 \; {\mu}^{3/2} } \int \! d^3\!x \,\mu^2\;  e^{\phi_0(\ve{x})}
\; \left({\cal G}(\ve{x},\ve{x}) \right)^2\;  {\cal G}(\ve{x},\ve{y}) + \ldots
 \nonumber \\
\end{eqnarray}

\be
g^{(2)}_{\chi}(\ve{x},\ve{y})= {\cal G}(\ve{x},\ve{y})
+ {g^2 \over 2 \; \mu} \int \! d^3 \! z \,\mu^2\;  e^{\phi_0(\ve{z})}\; 
{\cal G}(\ve{x},\ve{z})\; 
{\cal G}(\ve{z},\ve{z})\; 
{\cal G}(\ve{z},\ve{y})\, + \ldots
\ee

\noindent
The $g^{(n)}_{\chi}$ can be straightforwadly expressed in terms of 
Feynman diagrams.

\vskip 1.5cm
\noindent
$\displaystyle Z_{{\gr{x}}_0}[0]\,=\, $
\nopagebreak[4] 
\vskip -1.25cm
\nopagebreak[4]
\hskip 1cm
\epsfig{file=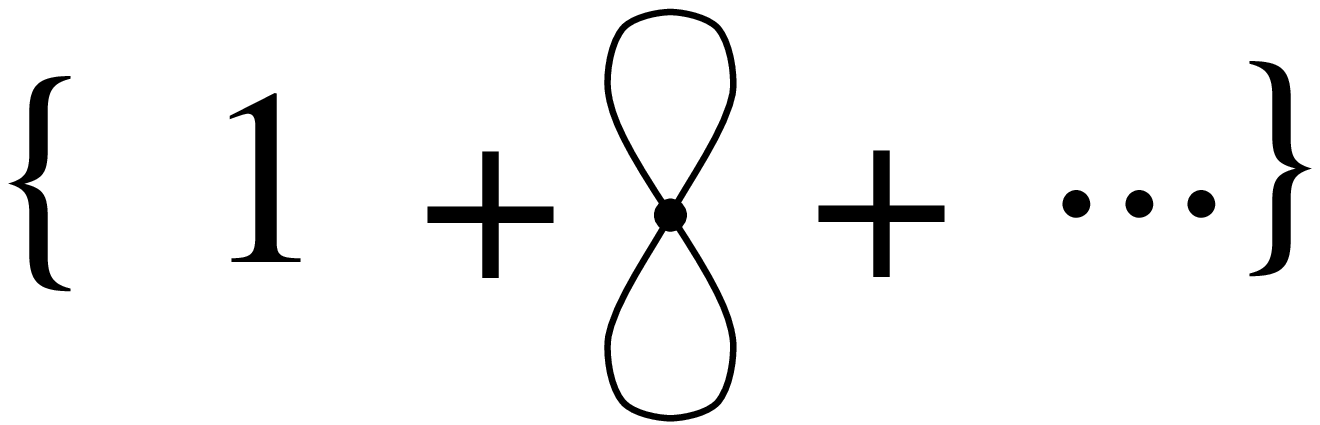,width=6cm}

\vskip 1cm
\noindent
$\displaystyle g^{(1)}_{\chi}\,=\, $
\nopagebreak[4]

\vskip -1.25cm
\hskip 1cm
\epsfig{file=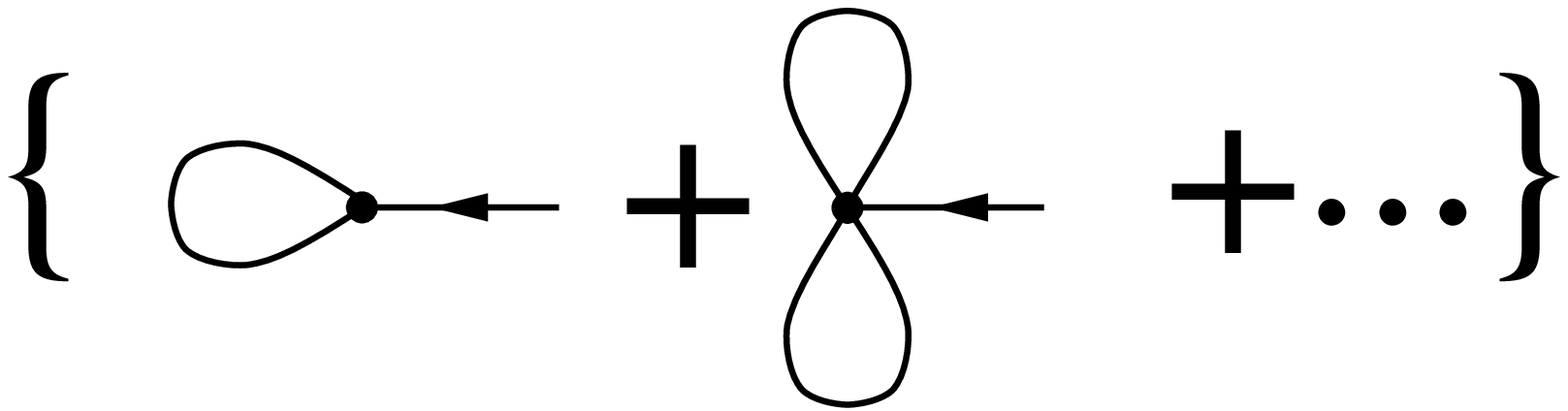,width=8cm}

\vskip 1cm
\noindent
$\displaystyle g^{(2)}_{\chi}\,=\, $

\vskip -1.25cm
\hskip 1cm
\epsfig{file=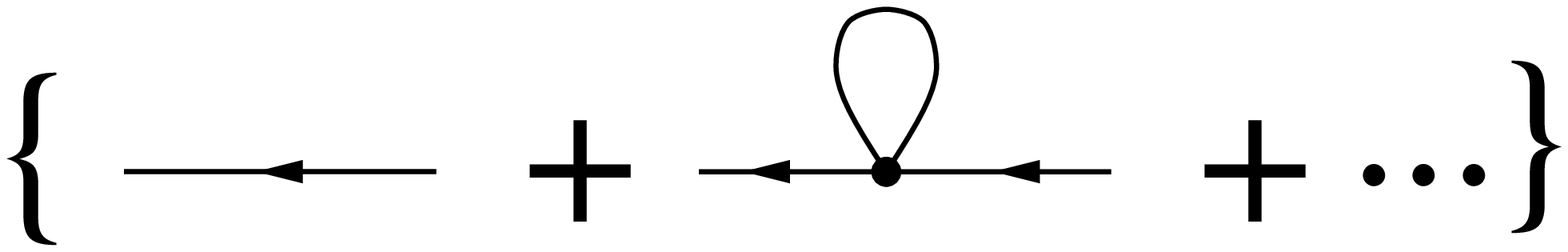,width=8cm}

\bigskip
\noindent
From these diagrams one can recover the previous 
integrals using the following Feynman rules:

\begin{itemize}
\item For each line: $ {\cal G}(\ve{x},\ve{y})$,
\item For a $n$-leg  vertex: $\displaystyle \quad + \left({g \over \sqrt{\mu}}
\right)^{n-2}
\int \!\!d^3\!{\gr{x}}\, \mu^2 \; e^{\phi_0(\ve{x})}$,
\item The diagram symmetry factor.
\end{itemize}

We consider now the connected correlation functions which are simpler
than the disconnected ones.

\subsubsection{Connected Green functions and vertex functions}

The generating functional of connected correlations is given by
\begin{equation}
W_{\gr{x}_0}[J]=\ln Z_{\gr{x}_0}[J].
\end{equation}
This is also a series in $J$ whose coefficients are the 
connected Green functions,

\be
W_{\gr{x}_0}^{(n)}(\gr{x}_1,\ldots,\gr{x}_n)= \left. { \delta^n
W_{\gr{x}_0}[J] \over \delta J(\gr{x}_1)
\ldots \delta J(\gr{x}_n) } \right|_{J=0}
\ee
The name connected arises because only {\em connected} Feynman diagrams
(in the topological sense) contribute to the perturbative 
expansion of the $W^{(n)}_{\gr{x}_0}$.

The expectation value of the field is given by the one-point function,
\be
\overline{\chi}(\gr{x})=
{ \delta W_{\gr{x}_0}[J] \over \delta J(\gr{x}) }.
\ee

The generating functional of the vertices (also called 1-particle irreducible) 
functions is defined by the Legendre transformation,

\be
\Gamma_{\gr{x}_0}[\overline{\chi}]+W_{\gr{x}_0}[J]=
\int  d^3\!\gr{x}  \; \overline{\chi} \; J 
\ee
Then, it is easy to derive the inverse  relation

\be
J(\gr{x})= {\delta \Gamma_{\gr{x}_0}[\overline{\chi}] 
\over \delta \overline{
\chi}(\gr{x}) }.
\ee
The 1-particle irreducible vertex functions are given as

\be
\Gamma_{\gr{x}_0}^{(n)}(\gr{x}_1,\ldots,\gr{x}_n)= \left.
{ \delta^n \Gamma_{\gr{x}_0}[\chi] \over
\delta \overline{\chi}(\gr{x}_1) \ldots \delta 
\overline{\chi}(\gr{x}_n) } \right|_{J=0},
\ee
Notice that the derivatives are taken at $J=0$, that is $\overline{\chi}=
W_{\gr{x}_0}^{(1)}$. So, the vertex functions are the coefficients of the
development of $\Gamma_{\gr{x}_0}$ around a non-zero field.

 The vertex functions are related
to the connected Green functions through,

\begin{eqnarray}
\Gamma_{\gr{x}_0}^{(1)}(\gr{x})&=&0,\nonumber\\
\Gamma_{\gr{x}_0}^{(2)}(\gr{x},\gr{y})&=& \left[ W_{\gr{x}_0}^{(2)}\right]^{-1}
\!\!(\gr{x},\gr{y}),\\
\Gamma_{\gr{x}_0}^{(3)}(\gr{x},\gr{y},\gr{z})&=&
 -\int \!\!d^3\!\gr{x}_1\;
d^3\!\gr{y}_1 \;d^3\!\gr{z}_1\,\,
\Gamma_{\gr{x}_0}^{(2)}(\gr{x},\gr{x}_1)\;
\Gamma_{\gr{x}_0}^{(2)}(\gr{y},\gr{y}_1)\;
\Gamma_{\gr{x}_0}^{(2)}(\gr{z},\gr{z}_1) \nonumber \\
&& \qquad \qquad \qquad \qquad \qquad \qquad\qquad
W_{\gr{x}_0}^{(3)}(\gr{x}_1,\gr{y}_1,\gr{z}_1)\; .\nonumber
\end{eqnarray}

Only one-particle irreducible Feynman diagrams contribute to the
 vertex functions, that is, diagrams which cannot be disconnected
by a single  cut of an internal line. In addition, external lines 
are   amputated to these diagrams. 

The knowledge of $\Gamma^{(2)}$ is crucial since it is the amputating operator.
It can be expressed as,

\be
\Gamma_{\gr{x}_0}^{(2)}(\gr{x},\gr{y})=
{\cal G}^{-1}(\gr{x},\gr{y}) - \Sigma(\gr{x},\gr{y}),
\ee
where $\Sigma(\gr{x},\gr{y})$ is the sum of one-particle irreducible
diagrams with two amputated external lines. In our theory,

\bigskip
 
\noindent
$\displaystyle
\Sigma(\gr{x},\gr{y})={ 1\over 2}\mu^2\;  e^{\phi_0(\gr{x})}\; {\cal
G}(\gr{x},\gr{x}) \; 
\delta(\gr{x}-\gr{y})\; {g^2 \over \mu}+ {1 \over 2}
\mu^2\;  e^{\phi_0(\gr{x})}\; \left({\cal G}(\gr{x},\gr{y})\right)^2\;  \mu^2
\; e^{\phi_0(\gr{y})} {g^2 \over \mu}
$
\smallskip
 
\noindent
$\hfill \displaystyle
+\,\,{3 \over 2}\,\,\mu^2 e^{\phi_0(\gr{x})}\left( {\cal G}(\gr{x},\gr{x})\right)^2
\delta(\gr{x}-\gr{y}) \,{g^4 \over \mu^2}+\ldots 
$

\bigskip
 \noindent
or, with diagrams,

\vskip 1.5cm
\noindent
$\displaystyle \Sigma({\gr{x}},\gr{y})\,=\, $
 
\vskip -1.4cm
\hskip 1.5cm
\epsfig{file=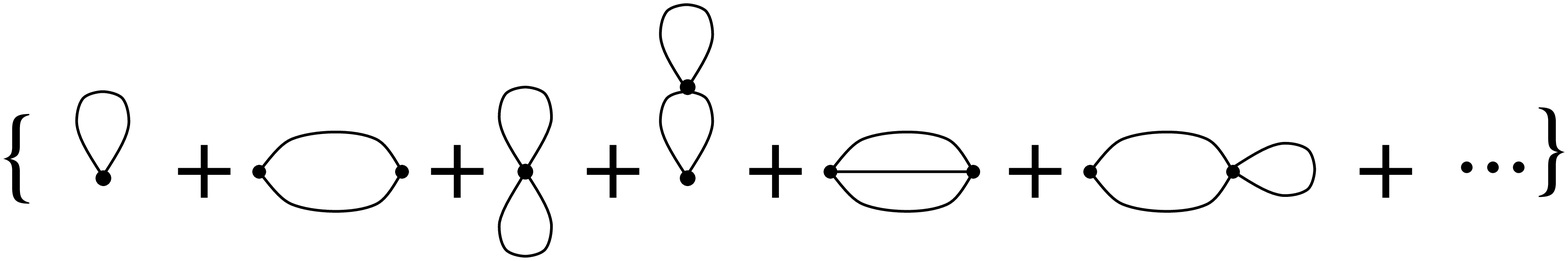,width=10cm}
\bigskip

\subsubsection{Computation of $\Gamma^{(2)}$ and 
$\Gamma^{(3)}$}

We consider in this section $\Gamma^{(2)}$ and $\Gamma^{(3)}$. These are
the relevant Green functions for renormalization. 
\vskip 1.5cm
\noindent
$\displaystyle \Gamma_{{\gr{x}}_0}^{(2)}=$
 
\vskip -0.75cm
\hskip 1cm
\epsfig{file=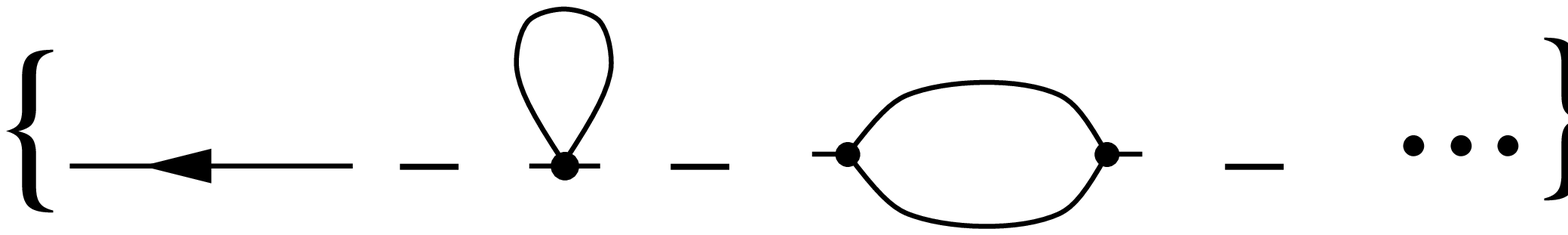,width=6cm}
 
\vskip 1.5cm
\noindent
$\displaystyle \Gamma_{{\gr{x}}_0}^{(3)}\,=\,-\,$
 
\vskip -0.75cm
\hskip 1cm
\epsfig{file=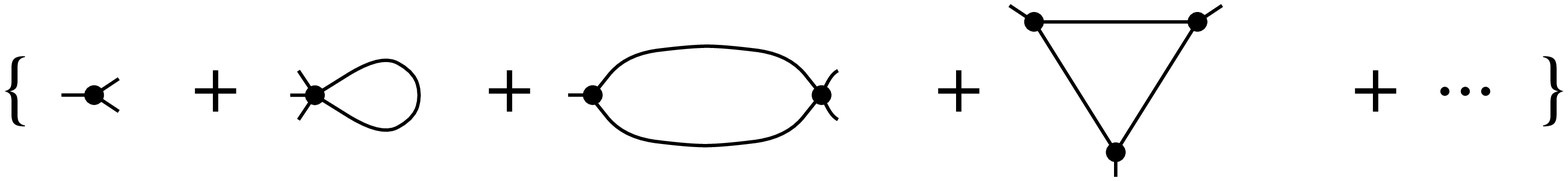,width=10cm}

\bigskip
\noindent
where the simple left arrow indicates the inverse propagator. Stumps are
meant to remind of the amputated external lines. From these diagrams we can
write the integrals applying our Feynman rules. We have finally to 
integrate over the collective coordinate ${\gr{x}}_0$.
 Moreover, choosing to integrate over all but one
of the vertex functions variables, yields more compact results
and  keep all the information about the renormalization.
We display here the results for the first diagrams, where we use the notation
$$
\tau=+\ln(R/\delta)  \;,
$$

As ultraviolet regularization we use a cutoff in the angular momentum
sum over $ l $. That is, in each propagator we sum up to a maximun $ L
$. Since $ l \sim  k r $, we use that the momentum cutoff is $ k \leq
1/ \delta $ and the size cutoff is $ r \leq R $. Therefore, we take
$$
l < L = R / \delta = e^{\tau} 
$$
in each propagator.

\bigskip
\noindent
\begin{tabular}{lll}
\epsfig{file=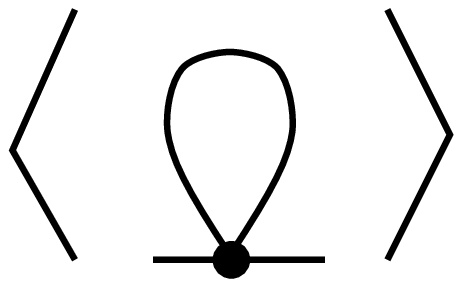,width=1.5cm}&$=$&
$\displaystyle  C_1 \, {g^2 \over \mu} \, \int d^3 \!\!\gr{x} \int \!\! 
d^3\gr{x}_0\; \delta(\gr{x}-\gr{y})\;{\cal G}(\ve{x},\ve{x})\;
\mu^2\; e^{\phi_0(\ve{x})}$\\
&&\\
&$=$& $\displaystyle
{g^2 \over \mu}\left[
\tau \; \sum_{l=0,l\neq 1}^{e^{\tau}} {{2l+1} \over
 2 \alpha_l } {\ch(\alpha_l \tau) \over
\sh(\alpha_l\tau) } \; \,-\, \sum_{l=0,l\neq 1}^{e^{\tau}}
{{2l+1} \over \alpha_l^2} \right]
$\\
\end{tabular}

\bigskip
\noindent
\begin{tabular}{lll}
\epsfig{file=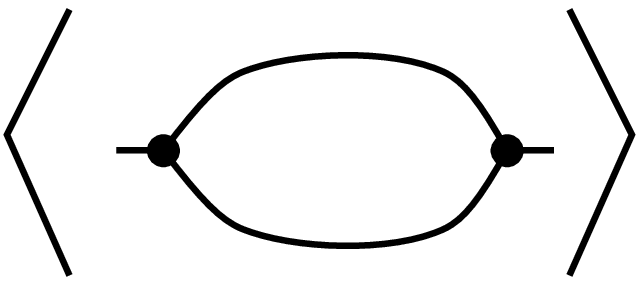,width=2cm}&$=$&
$
\displaystyle  
+C_2 \,{g^2 \over \mu} \, \int d^3 \!\!\gr{x} \int \!\! d^3\gr{x}_0\; 
{\cal G}^2(\ve{x},\ve{y})\; 
\mu^2\;  e^{\phi_0(\ve{x})}\; \mu^2\; e^{\phi_0(\ve{y})}
$\\
&&\\
&$=$& $\displaystyle
 \,{g^2 \over \mu} \, \left[{{\tau^2} \over 2} \sum_{l=0,l\neq 1}^{e^{\tau}}
{ {2l+1} \over \alpha_l^2 \sh^2( \alpha_l \tau )} 
+{{\tau} \over 2} \sum_{l=0,l\neq 1}^{e^{\tau}}
{ (2l+1) \ch (\alpha_l \tau ) \over
 \alpha_l^3 \sh( \alpha_l \tau )}
- \sum_{l=0,l\neq 1}^{ e^{\tau}} { {2l+1} \over \alpha_l^4 } \right]
 $\\
\end{tabular}

\bigskip
\noindent
\begin{tabular}{lll}
\epsfig{file=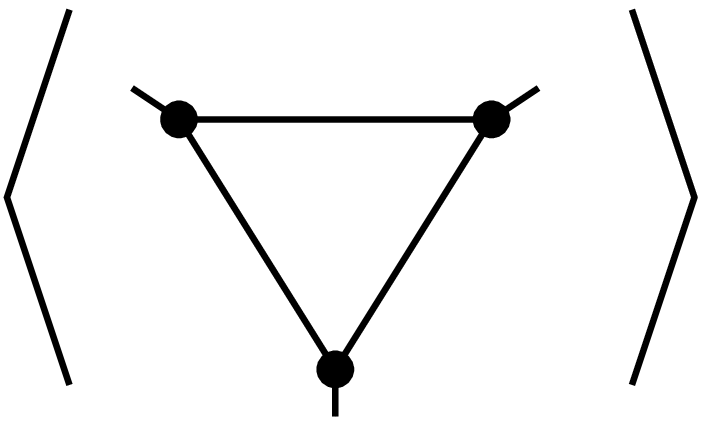,width=1.5cm}&$=$&
$\displaystyle  C_5 \left({g \over \sqrt{\mu}}\right)^3 
\int \!\!d^3 \gr{y}\int \!\!d^3 \gr{x} \int \!\! d^3\gr{x}_0 
{\cal G}(\ve{x},\ve{y})\;  {\cal G}(\ve{y},\ve{z}) \; {\cal G}(\ve{z},\ve{x})
$\\
&&\\
&&$\displaystyle \qquad \qquad \times \,\,
\mu^2\;  e^{\phi_0(\ve{x})}\; 
\mu^2\;  e^{\phi_0(\ve{y})}\; \mu^2\;  e^{\phi_0(\ve{z})}$
\\
&&\\
&$=$& $\displaystyle
\left({g \over \sqrt{\mu}}\right)^3 \left[\tau^3  \sum_{l=0,l\neq
1}^{e^{\tau}} 
{ (2l+1) \ch(\alpha_l \tau) \over  \alpha_l^3\sh^3( \alpha_l \tau)} 
+ {{3 \tau^2}\over 2} \sum_{l=0,l\neq 1}^{e^{\tau}}
{ {2l+1} \over  \alpha_l^4 \sh^2( \alpha_l \tau )} \right.
 \qquad$ \\
&&\\
&& \hfill $\displaystyle \left.
+ {{3\tau} \over 2} \sum_{l=0,l\neq 1}^{e^{\tau}}
{ (2l+1) \ch(\alpha_l \tau) \over \alpha_l^5
\sh( \alpha_l \tau )} 
\,-\,4 \sum_{l=0,l\neq 1}^{e^{\tau}}
{ {2l+1} \over  \alpha_l ^6 } \right]
$ \\
\end{tabular}
\bigskip

The $C_i$ factors are the symmetry factors of the diagrams. $C_3$ is the 
factor for the tadpole with three amputated external lines instead of two, and
$C_4$ is the eye-like diagram with two amputated lines on one side and one on
the other. Their values are, 
$$
C_1=C_2=C_3={1 \over 2} \; , \; C_4={3 \over 2}  \; \mbox{and}\;
C_5=1 \; .
$$

The only averaged diagrams left to compute, at this order of approximation, are
the most simple ones, the vertex and the inverse propagator. 
They yield

\be
\int_{\cal V} \!\! d^3\gr{x}_0 \int_{\cal V} \!\! d^3 \gr{x} \; 
{\cal G}^{-1}(\ve{x}, \ve{y}) \,=\, -8 \pi (R- \delta)\; ,
\ee
\be
\int_{\cal V} \!\! d^3\gr{x}_0 \int_{\cal V} \!\! d^3 \gr{x} 
\int_{\cal V} \!\! d^3 \gr{y}\;  \delta(\gr{x}-\gr{y}) \delta(\gr{y}-\gr{z})
\; {g \over \sqrt{\mu} } \mu^2 e^{\phi_0(\ve{x})} \,=\, 8\pi(R-\delta)\,
{g \over \sqrt{\mu}},
\ee
where ${\cal V}$ is  the volume of the region between the spheres with
radii $\delta$ and $R$. The expansion  of the  zero-momentum  vertex functions
take the form,

\begin{eqnarray} \label{ecuC}
\langle \Gamma^{(2)} \rangle\, &=& \,- {8 \pi R  \over {\cal N}}
\left\{ 1- e^{-\tau} + \left({g^2 \over \mu R}\right){1 \over 8 \pi} \left[
+\tau^2  \sum_{l=0,l\neq 1}^{e^{\tau}}
{ {2l+1} \over 2\alpha_l^2 \sh^2( \alpha_l \tau )} 
\right. \right. \cr \cr
 \qquad \qquad \qquad\,&+&\, \left. \left. \tau  \sum_{l=0,l\neq
1}^{e^{\tau}} {{2l+1} \over 2 \alpha_l^3 } 
{\ch(\alpha_l \tau) \over \sh(\alpha_l\tau) }
(\alpha_l^2 +1) - \sum_{l=0,l\neq 1}^{e^{\tau}} { {2l+1} \over \alpha_l^4 }
(1+{\alpha_l^2 \over 2}) \right] \,\,\, \right\}  \cr \cr
&& \nonumber \\
\langle \Gamma^{(3)} \rangle\, &=&\, -{8 \pi R^{3 \over 2} \over {\cal N}}
{g \over \sqrt{\mu R}}
\left\{ 1- e^{-\tau} + \left( { g^2 \over \mu R} \right)  { 1 \over 8 \pi} 
\left[ \,\tau^3  \sum_{l=0,l\neq 1}^{e^{\tau}}
{ (2l+1) \ch(\alpha_l \tau) \over  \alpha_l^3 \sh^3( \alpha_l \tau)}
\right.\right.\cr \cr
&+&\tau^2  \sum_{l=0,l\neq 1}^{e^{\tau}}
{ 3(2l+1) \over  2\alpha_l^4 \sh^2( \alpha_l \tau )}
(1+\alpha_l^2) 
+ \tau  \sum_{l=0,l\neq 1}^{e^{\tau}}{ (2l+1) \ch(\alpha_l \tau) \over
2\alpha_l^5 \sh( \alpha_l \tau )}
(3+3\alpha_l^2+\alpha_l^4)  
\cr \cr &-& \left. \left. \sum_{l=0,l\neq 1}^{e^{\tau}}
{ 2l+1 \over 2 \alpha_l ^6 } (8+6\,\alpha_l^2+\alpha_l^4 )
\right] \right\}
\end{eqnarray}

We have now all the quantities to renormalize the coupling
constant at one loop order. It must be noticed that the perturbative
expansion turns out to be in powers of the effective coupling
\be\label{defla}
\lambda \equiv {g^2 \over \mu R}  \; \; .
\ee
This fact will be very important for the behaviour of the physical quantities.

\subsubsection{Zero-momentum value of the $\Gamma^{(n)}$}
 
Let us investigate the $n$ points vertex functions. We can write $\Gamma^{(n)}
_{\gr{x}_0}$ at first loop order as a sum of rather simple diagrams,
 
\vskip 1.5cm
\noindent
$\displaystyle \Gamma_{\gr{x}_0}^{(n)}\,=\,-$
\nopagebreak[4] 
\vskip -0.90cm
\hskip 1cm
\nopagebreak[4]
\epsfig{file=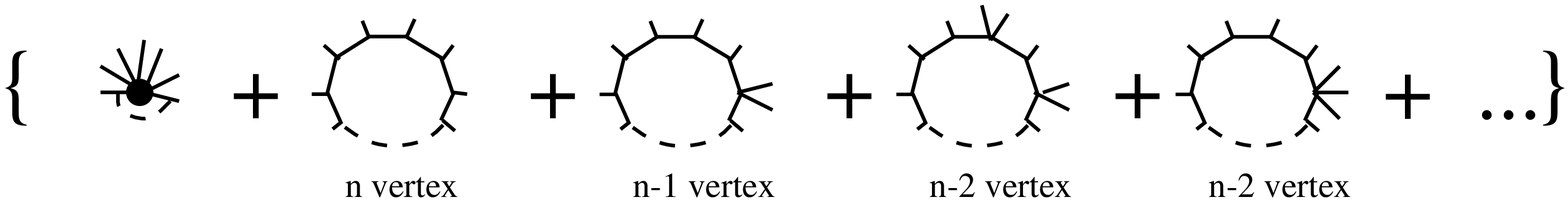,width=11cm}
 
\bigskip
\noindent
In this summation we should also include all the diagrams deduced from those
drawn here by topological permutations of the vertex within a given diagram.
This would create families of diagrams, with each member of a single family
containing $k_i$ $i$-legged vertex. However, all members of a given family have
the same zero-mementa value (this can be checked on the corresponding
integrals). 

It is enough for renormalization to compute the diagrams at zero-momenta.
Each diagram drawn above will stand for the whole family and will be
assigned the sum $C_{k_3,\ldots,k_{n+2}}$ of the symmetry coefficients of the
members of the family. The averaged zero-momenta value of
a representative diagram will be noted $\langle \{k_3,\ldots,k_{n+2}\}\rangle$.
We proceed now to compute this value.
 
Let us call $ v=\sum_i  k_i $ the number of vertex in a diagram. Then,
 
\be \label{vernum}
\sum_j j k_j \,=\, n +2v \;.
\ee
We can now recast the diagrams into integrals using Feymann rules and
eq.(\ref{vernum}),

\begin{eqnarray}  \label{basicdiag}
\langle \{k_3,\ldots,k_{n+2}\}\rangle &=&
C_{k_3,k_4,\ldots,k_{n+2}}\, \left( {g \over \sqrt{\mu}} \right)^n
\int\!\! d^3\!\gr{x}_0 d^3\!\gr{x}_1 \ldots d^3\!\gr{x}_{v-1}
{\cal G}(\ve{x_1},\ve{x_2}) \ldots
{\cal G}(\ve{x_v},\ve{x_1}) \nonumber \\
&& \nonumber \\
&& \qquad \qquad \qquad \qquad \qquad \qquad \qquad \qquad \qquad
\times \mu^2 e^{\phi(\ve{x_1})}
\ldots \mu^2 e^{\phi(\gr{x_v})}\;.  \nonumber\\
\end{eqnarray}
The sum $C_{k_3,k_4,\ldots,k_{n+2}}$ is actually easier to compute than each
single symmetry coefficient. It can be written as:
 
\be
C_{k_3,k_4,\ldots,k_{n+2}} \,=\, {n! v! \over 2v \,k_3!\, k_4! \,\ldots\,
k_{n+2}! } \prod_{j=3}^{n+2} \left( {1 \over (j-2)! } \right)^{k_j}.
\ee
Inserting the explicit propagator in eq.(\ref{basicdiag}) and integrating over
the angular variables yields:

\begin{eqnarray}
\langle \{ k_3,k_4,\ldots,k_{n+2} \} \rangle &\,=\,&
C_{k_3,k_4,\ldots,k_{n+2}} 2^v \sum_{l=0,l\neq 1}^{e^{\tau}} \int \!\!
dr_1 \ldots dr_v g_l(r_1,r_2) \ldots g_l(r_v,r_1)\nonumber\\ 
&&\nonumber\\
&=& C_{k_3,k_4,\ldots,k_{n+2}} 2^v
\sum_{l=0,l\neq 1}^{e^{\tau}}
{\omega_v(\alpha_l \tau) \over \alpha_l^v \sh^v(-\alpha_l \tau)}\,
\\ \nonumber
\end{eqnarray} 
where
 
\be
 \omega_v(x)\,=\, v!\,\int_0^{x}\!\!\!\! dr_1 \int_0^{r_1}\!\!\!\! dr_2 \ldots
\int_0^{r_{v-1}} \!\!\!\! dr_v \sh^2(r_1+x) \prod_{k=2}^{v-1}
 \sh(r_k)\sh(r_k+x) \,\,\, \sh^2(r_v)\;.
\ee

The coefficients $\omega_v$ can be directly calculated with formal
calculus softwares 
(we have done it for $v \leq 5$). In addition,  it can be established that they
obey the following recurrence formula: 

\begin{eqnarray}
\omega_v(x) &=& {2^v (\ch(x)-1)^v \over x^v} \left\{ (-f_1(x))^v \, f_v(x)\,+\,
\sum_{i=1}^{v-1}\left[(-1)^{i+1} \left( {\omega_{v-i}(x) \, x^{v-i} \over
2^{v-i} \, (\ch(x)-1)^{v-i} }  \right. \right. \right. \cr \cr
&-& (-f_1(x))^{v-i} \, f_{v-i}(x) \Bigg) f_{i+1}(x)
\, f_1^{i-1}(x) \Bigg] - (-1)^n f_{v+1}(x) \, f_1^{v-1}(x) \Bigg\} \nonumber
\end{eqnarray}
where,
\begin{eqnarray}
f_1(x) &=& { {x \over 2} \over \tanh({x \over 2}) } \, ,\nonumber \\
&& \nonumber \\
f_{n+1}(x) &=& f_n(x) - {x \over 2 n} f_n^{\prime}(x) \;. \nonumber 
\end{eqnarray}
This recurrence relation allows us to compute the $\Gamma^{(n)}$
without performing integrations.
\be
\langle \Gamma^{(n)} \rangle \,=\, \sum_{\{k_i, \sum_j (j-1) k_j  =v\}}
C_{k_3,\ldots,k_{n+2}} 2^v \sum_{l=0,l\neq 1}^{e^{\tau}} {\omega_v(\alpha_l  
\tau) \over \alpha_l^v \sh^v(-\alpha_l \tau)}
\ee

However, as far as renormalization is concerned, we need
only the leading order of the $\Gamma^{(n)}$ in $\tau$. We can first establish
that:

\begin{eqnarray}
\langle \{ k_3,\ldots,k_{n+2} \} \rangle &=& C_{k_3,\ldots,k_{n+2}}\,
\left[ \tau^v \, \sum_{l=0,l\neq 1}^{e^\tau} {(2l+1) \over
\alpha_l^v} {\ch^{v-2}(\alpha_l \tau) \over \sh^{v}(\alpha_l \tau)} \right. \,
\cr \cr 
&+& \left. \tau^{v-1} \,\,
{v \over 2}\sum_{l=0,l\neq 1}^{e^\tau} {(2l+1) \over
\alpha_l^{v+1}} {\ch^{v-3}(\alpha_l \tau) \over \sh^{v-1}(\alpha_l \tau)}
+ {\cal O}(\tau^{v-2}) \right]  \nonumber\\
\end{eqnarray}
Accordingly, only the diagrams with $n$ and $n-1$ vertex contribute to the
two leading orders of $\Gamma^{(n)}$ in $\tau$. Since $ \displaystyle
C_{n,0,\ldots,0}= { (n-1)! \over 2}$ and $ \displaystyle
C_{n-2,1,0,\ldots,0}={n! \over 4}$, we find the explicit formula,

\begin{eqnarray}\label{Gaman}
\langle \Gamma^{(n)} \rangle & = &
-{8 \pi R^{{n \over 2}} \over {\cal N}} \left( {g \over \sqrt{\mu
R}}\right)^{n-2}  
\left\{ 1- e^{-\tau} + \left( {g^2 \over \mu R} \right)
{1 \over 8 \pi} \left[ {(n-1)! \over 2} \; \tau^n \!
\sum_{l=0,l\neq 1}^{e^\tau} {(2l+1) \over
\alpha_l^n} {\ch^{n-2}(\alpha_l \tau) \over \sh^{n}(\alpha_l \tau)}
\right. \right. \cr \cr
&+&\left. \left. {n! \over 4} \; \tau^{n-1} \left(\sum_{l=0,l\neq 1}^{e^\tau}
{(2l+1) \over \alpha_l^{n+1}}
{\ch^{n-3}(\alpha_l \tau) \over \sh^{n-1}(\alpha_l \tau)}(1+\alpha_l^2)
\right) + {\cal O}(\tau^{n-2}) \right] \right\} \nonumber 
\end{eqnarray}

Before we turn to renormalization let us study the density correlation
functions. 

\subsection{Density correlations}

The two basic local fields here are the gravitational potential $
\phi(x) $ and the particle density defined as in eq.(\ref{tabla}),
$$
\rho({\vec x}) =  -{1 \over {T_{eff}}}\;\nabla^2 \phi({\vec x}) \; .
$$
The correlation functions of the local density  $ \rho({\vec x})  $
and their long-range behaviour are particularly important.  

Let us  introduce a source term in the partition function in order to
compute density correlations.

\be
Z_{\rho}[\sigma] = \int \!\! D\phi \; e^{-S[\phi]+\int d^3x \;
\rho({\vec x})\, \sigma({\vec x}) } \; .
\ee
We then define de density correlation functions as

\be
\left.
G^{(n)}_{\rho}(x_1,\ldots,x_n) \,=\, {1 \over Z_{\rho}[0]}
{ \delta^n Z_{\rho}[\sigma] \over \delta\sigma(x_1) \ldots \delta\sigma(x_n)}
\right|_{\sigma=0}\; .
\ee

The generating functionnal of the 
connected correlation function of the density field is defined as
usual as

\be
W_{\rho}[\sigma]= \ln Z_{\rho}[\sigma] \; .
\ee

We finally define the generating functionnal of the vertex correlation 
functions of the density field:

\be
\Gamma_{\rho}\left[\overline{\rho}\right] + W_{\rho}[\sigma]\,=\,
\int d^3x \;  {\overline{\rho}}(x)\,\sigma(x)  \; ,
\ee
\be 
{\overline{\rho}}(x)\,=\, { \delta W_{\rho}[\sigma] \over \delta
\sigma(x) } \; . 
\ee

At tree level the two point density correlator takes the form
\begin{eqnarray}
<\rho({\vec x})\, \rho({\vec y}) > &=&\left({1 \over {T_{eff}}}\right)^2
\int {d^3 \gr{x}_0 \over {\cal N} } < \nabla^2_x \chi({\vec x}-{\vec
x}_0)\, \nabla^2_x \chi({\vec y}-{\vec x}_0)>\cr \cr
&=&{4 \over {\cal N}\; {T_{eff}}^2}\int d^3 \gr{x}_0 \;
{ {\cal G}({\vec x}-{\vec x}_0 , {\vec y}-{\vec x}_0  ) 
\over |{\vec x}-{\vec x}_0|^2 \;  |{\vec y}-{\vec x}_0|^2 }\quad ,
\end{eqnarray}
where we used eq.(\ref{laplaG}) in the course of the calculation.

Using that ${\cal G}(\hbox{\bf x},\hbox{\bf y}) $ scales as $ 1
/\sqrt{|{\vec x}||{\vec y}|} $, 
we see that this correlator scales as $ 1/|{\vec x}-{\vec y}|^2  $.

\section{ Renormalization}
 
As usual, the renormalized vertex functions are defined as follows:
\be
\Gamma^{(n)}_{\cal R}(x_1,\ldots,x_n, g_R) \equiv Z^{n/2}\;
\Gamma^{(n)}(x_1,\ldots,x_n, g)\; .
\ee
where $ Z $ stands for the wave function renormalization, $ g $ is the
bare coupling and $ g_R $ the renormalized one. 

The renormalizations are defined such that the two and three points
functions take their tree level values at zero external momenta.
In our case, from eq.(\ref{ecuC})  the renormalization prescriptions
take the form   

\begin{eqnarray}
\displaystyle\label{presren}
\langle \Gamma^{(2)}_{\cal R}  \rangle \,&=&\, - { 8 \pi R\over {\cal N}}
(1-e^{-\tau})  \;,\cr \cr
\langle \Gamma^{(3)}_{\cal R} \rangle \,&=&\,  {8 \pi R^{3 \over
2}\over {\cal N}} (1-e^{-\tau}) 
\left( {g_{\cal R} \over \sqrt{\mu R}} \right) \; . 
\end{eqnarray}
The bare vertex functions  at the one-loop level are given by
eq.(\ref{ecuC}) and can be summarized as follows,

\begin{eqnarray}\label{gama2}
\langle \Gamma^{(2)} \rangle\, &=&\,- {8 \pi R \over {\cal N}}
 (1-e^{-\tau}) 
\left[ 1 + \left( { g^2 \over \mu R} \right)  { C(\tau) \over 8 \pi}  \right],
\cr \cr
\label{gama3}
\langle \Gamma^{(3)} \rangle\, &=&\, { 8 \pi R^{3 \over 2} \over {\cal N}}
(1-e^{-\tau})  \left( {g \over \sqrt{\mu R}} \right)
\left[ 1 + \left( { g^2 \over \mu R} \right)  { D(\tau) \over 8 \pi}
\right]\; .
\end{eqnarray}
where $C(\tau)$ and $D(\tau)$ can be read off from eqs.(\ref{ecuC}). 

The bare coupling $g$ and the wave function renormalization can be
expressed as, 
\begin{eqnarray}\label{reng}
g_R\,&=&\, g \left[ 1 + \left({g^2 \over \mu R} \right)\; {1 \over 8 \pi}\;
h(\tau) 
+ {\cal O}\left( {g^2 \over \mu R}\right)^2\right]\; ,
\cr \cr
Z \, &=& \, 1 + \left({g^2 \over \mu R} \right) {1 \over 8 \pi}\; B(\tau) +
{\cal O}\left( {g^2 \over \mu R}\right)^2 \; .
\end{eqnarray}
$h(\tau)$ and $B(\tau)$ follow by inserting eqs.(\ref{gama2})-(\ref{reng}) 
into the renormalization prescriptions eq.(\ref{presren}) with the result,
\be
h(\tau)\,=\, D(\tau)-{3 \over 2}C(\tau)\quad , \quad 
B(\tau)\,=\,-C(\tau)\; .
\ee
We can thus write $ Z(\tau) $ and $ g_R $ up to the first order in ${g^2
\over \mu R}$:
\begin{eqnarray}
Z(\tau) &=& 1 + \left( {g^2 \over \mu R} \right){1 \over 8 \pi} \left[
\,-\, \tau^2  \sum_{l=0,l\neq 1}^{e^{\tau}}{{2l+1}\over2\; 
\alpha_l^4 \sh^2(\alpha_l \tau )}
\,-\,\tau  \sum_{l=0,l\neq 1}^{e^{\tau}}
{ (2l+1) \ch(\alpha_l \tau) \over 2\alpha_l^5 \sh( \alpha_l \tau )}
(1-\alpha_l^2) \right. \nonumber \\
&& \nonumber \\
&& \left. \qquad \qquad \qquad \qquad \qquad \qquad \qquad \qquad
\,+\, \sum_{l=0,l\neq 1}^{e^{\tau}}
{ {2l+1} \over  \alpha_l ^6 } (1+{\alpha_l^2 \over 2} )+ {\cal O}\left({g^2
\over \mu R}\right) \right] \nonumber \\ 
\end{eqnarray}
 
\begin{eqnarray}\label{gR}
g_{\cal R}&=& g\, \left\{ 1+\left( {g^2 \over \mu R} \right){1 \over 8 \pi}
\left[\,\tau^3  \sum_{l=0,l\neq 1}^{e^{\tau}}
{ (2l+1) \ch(\alpha_l \tau) \over  \alpha_l^3 \sh^3( \alpha_l \tau )}
\,+\,\tau^2 \sum_{l=0,l\neq 1}^{e^{\tau}}{(2l+1) \over  
2\alpha_l^4 \sh^2( \alpha_l \tau )} (3+{3 \over 2}\alpha_l^2)
\right. \right. \cr \cr
&+& \tau  \sum_{l=0,l\neq 1}^{e^{\tau}}
{ (2l+1) \ch(\alpha_l \tau) \over
2\alpha_l^5 \sh( \alpha_l \tau )}
(3+{3 \over 2}\alpha_l^2-{1 \over 2}\alpha_l^4)  
\left. \left.
\,-\, \sum_{l=0,l\neq 1}^{e^{\tau}}
{ {2l+1} \over  \alpha_l ^6 } (4+{3 \over 2}\alpha_l^2-{1 \over 4}\alpha_l^4 )
\,+\, {\cal O}\left({g^2 \over \mu R}\right)
\right] \,\,\,\right\} \nonumber \\
\end{eqnarray}
 
\subsection{The renormalization group equations}

\subsubsection{Derivation  of the renormalization group equations}

The renormalization group equations for the vertex functions can be
derived as usual, by requiring the renormalized vertices to be
independent of the  scale $ \tau $ except for the {\bf explicit}
dependence on the size of the system here arising from the position of
the boundary.

We therefore apply on  the bare correlators the differential operator
\be {\partial \over \partial \tau}= \frac12 \left(R 
{\partial \over \partial R}-\delta{\partial \over \partial \delta}
\right)
\ee
keeping the bare coupling $ \lambda $ defined in eq.(\ref{defla}) fixed.

We have,
$$\left.
{\partial \over \partial
\tau}\right|_{\lambda}\Gamma^{(n)}({\gr{x}}_i,\lambda,\tau)\,
=\,\Delta\Gamma^{(n)} \;. 
$$
where the insertion $ \Delta $ arises through  the $ \tau
$ derivative acting on the bounds of the action:

$$
\Delta \equiv -{1 \over 2\, \teff}\int \!\!d \Omega_{\gr{u}}\;
\left(R^3 {\cal L}(R \gr{u}) + \delta^3 {\cal L}(\delta \gr{u} ) \right) \;. 
$$
Since we have chosen zero boundary conditions, only the insertion of
field derivatives contribute to the correlators $\Gamma^{(n)}$.
We can then set
$$
 \Delta = -{1 \over 4\,\teff}\int \left( R^3 (\nabla \chi |_{R \gr{u}})^2 +
\delta^3 ( \nabla \chi |_{\delta \gr{u}} )^2 \right ) 
d \Omega_{\gr{u}} \;. 
$$
Since the boundaries are at finite
distances which scale with $ \tau $,  their effect must be taken into
account through the $\Delta$ insertion. However, on physical grounds
the boundary effects through the insertion $\Delta$  are expected to  be
subdominant. In addition, we show in appendix B that the insertion has indeed
negligeable effects.  

\medskip

By using the relation between renormalized and bare vertex functions
we get,

\be
\displaystyle
\left[ {\partial \over \partial \tau} + \beta(\lambda_{\cal R}) {\partial \over
\partial \lambda_{\cal R}} - {n-3 \over 2}\gamma(\lambda_{\cal R}) \right] 
\Gamma_{\cal R}^{(n)}({\gr{x}}_i,\lambda_{\cal
R},\tau)\,=\,\Delta\Gamma_{\cal R}^{(n)} \;. 
\ee
where  we introduced the renormalized effective coupling
$$
 \lambda_{\cal R}\equiv
{g_{\cal R}^2 \over \mu R} \; .
$$

The functions $\beta$ and $\gamma$ are defined by:

\be \label{funbeta}
\beta(\lambda_{\cal R})\,=\, \left. {\partial \lambda_{\cal R} \over
\partial \tau} \right|_{\lambda}
\ee

\be
\gamma(\lambda_{\cal R})\,=\,  \left. {\partial \ln Z \over \partial \tau}
\right|_{\lambda}.
\ee

We find from eq.(\ref{Gaman}) that
$$
\gamma(\lambda_{\cal R})\,=\, 1 + {\cal O}(\lambda_{\cal R})\; .
$$

We now proceed to compute the RG flow of the effective coupling $
\lambda(\tau) $.

\subsubsection{RG flow of the effective coupling $\lambda(\tau) $}

The expansion of the effective coupling constant $\lambda$ easily
follows from the expansion of  $g$, eq.(\ref{gR}), 

\be
\lambda\,=\, \lambda_{\cal R}\, \left[ 1 \,-\, 
\lambda_{\cal R}{1 \over 4 \pi} h(\tau) \right].
\ee
Now, using eq.(\ref{funbeta}), we have:

$$ \displaystyle
\beta(\lambda, \tau) \,=\, -{ \lambda^2
 \over 4 \pi}\, {d h(\tau) \over d\tau}
$$
where $  h(\tau) $ is given by eq.(\ref{reng}) and (\ref{gR})

\begin{eqnarray}\label{hache}
h(\tau)&=&
\tau^3 \;  \sum_{l=0,l\neq 1}^{e^{\tau}}
{ (2l+1) \ch(\alpha_l \tau) \over  \alpha_l^3
\sh^3( \alpha_l \tau )} 
+\tau^2 \; \sum_{l=0,l\neq 1}^{e^{\tau}}
{ (2l+1) \over  2\alpha_l^4 \sh^2( \alpha_l \tau )}
\left(3+{3\over2}\alpha_l^2 \right) \cr\cr
&+& {\tau \over 4}\; \sum_{l=0,l\neq 1}^{e^{\tau}}
{ (2l+1) \ch(\alpha_l \tau) \over\alpha_l^5 \sh( \alpha_l \tau )}
\left(6+3 \alpha_l^2- \alpha_l^4 \right) 
-\frac14 \sum_{l=0,l\neq 1}^{e^{\tau}}
{ 2l+1 \over  \alpha_l ^6 } \left(16+6\alpha_l^2-\alpha_l^4 \right) \; .
\end{eqnarray}

The renormalization group flow of $ \lambda $ is determined by $
 \beta(\lambda, \tau) $ as usual by
$$
{ d \lambda \over d \tau} =  \beta(\lambda, \tau)
$$
Notice that contrary to the usual cases, there is a non-trivial and
explicit dependence on the scale $ \tau $ in $  h(\tau) $,  and hence in
$  \beta(\lambda, \tau)$. We get,

\be
{d\lambda \over \lambda^2} \,=\, -{1 \over 4 \pi}\; h^{\prime}(\tau)\;
d\tau.
\ee

Integrating this expression from $ \tau = \tau_i $ to $ \tau $, we
find for  $\lambda(\tau)$ 

\be \label{lambeff}
\displaystyle
\lambda(\tau) \, = \, { \lambda_i \over 1 + {\lambda_i \over 4 \pi}\left[
h(\tau) - h(\tau_i) \right] } \; ,
\ee
 
where $ \lambda_i \equiv \lambda(\tau_i) $.

\bigskip
\noindent
We proceed now  to analyse the nontrivial dependence of
the effective coupling constant $ \lambda(\tau) $ on the scale $ \tau $.

Let us define the quantities,
$$
\sigma_n = \sum_{l=2}^{e^\tau} { 2l+1 \over \alpha_l^n} \; .
$$
We will use the $ \sigma_n $ for $ 1 \leq n \leq 6 $:

\begin{itemize}
\item
$\displaystyle
\sigma_1\,=\, 
2(\zeta(0)-1)+\sum_{l=2}^{e^\tau} \left({ 2l+1 \over
\alpha_l}-2 \right) \,=\, -1.874677\,+\,O(e^{-\tau})
$
\item
$\displaystyle
\sigma_2\,=\, \,2\tau\, -0.7871527 \,+\, 2 \; e^{-\tau} \,+\,
 {\cal O}\left(e^{-2\tau}\right) 
$
\item
$\displaystyle \sigma_3
 \,=\, 1.314771\,+\,2 \; e^{-\tau}\,+\, {\cal O}\left(e^{-2\tau}\right) 
$
\item
$\displaystyle
\sigma_4\,=\, 0.4134319\,+\, e^{-2\tau}\,+\,{\cal
O}\left(e^{-3\tau}\right)  
$
\item
$\displaystyle
\sigma_5\,=\, 0.16731628 \,+\, \frac23\, e^{-3\tau} +\,
 {\cal O}\left(e^{-4\tau}\right) 
$
\item
$\displaystyle
\sigma_6\,=\, 0.07402368\,+\,{e^{-4\tau} \over 2}\,+\, {\cal
O}\left(e^{-5\tau}\right) 
$
\end{itemize}
Notice that only $ \sigma_2 $ grows logarithmically with the cutoffs. 

\noindent
This enables us to reduce $h(\tau)$ to the following  form:
\noindent
\begin{eqnarray}
\displaystyle
h(\tau) &=&- {\tau^3 \over \sin^3\left(
{\sqrt7 \over 2} \tau \right) } {8 \over 7\sqrt7 } \cos\left(
{\sqrt7 \over 2} \tau \right)- {3 \over 49}{\tau^2 \over \sin^2\left(
{\sqrt7 \over 2} \tau  \right)}\cr \cr
&+& {37 \over 98 \sqrt7}\; {\tau \over \sin\left( {\sqrt7 \over 2}
\tau \right) } 
\cos\left( {\sqrt7 \over 2} \tau \right)
\,+\,\tau\left({3 \over 2}\sigma_5 +{3 \over 4}\sigma_3-
{1 \over 4}\sigma_1+1 \right) \cr \cr
&-&4\sigma_6 -{3 \over 2}\sigma_4 + {1 \over 2}(\sigma_2 - 2 \tau)
\,-\,{10 \over 7^3}\; , 
\end{eqnarray}
or with the numerical values,
\begin{eqnarray}
\displaystyle
h(\tau) &=&-\, {\tau^3\over \sin^3\left(
{\sqrt7 \over 2} \tau \right) } {8 \over 7\sqrt7 } \cos\left(
{\sqrt7 \over 2} \tau \right) -{\tau^2 \over \sin^2\left(
{\sqrt7 \over 2} \tau  \right)} {3 \over 49} \cr \cr
&+&{\tau \over \sin\left( {\sqrt7 \over 2} \tau \right) }
\cos\left( {\sqrt7 \over 2} \tau \right) {37 \over 98 \sqrt7}
\,-\,0.70\,\tau-0.55+{\cal O}\left(\tau e^{-\tau}\right) 
\end{eqnarray}

The effective coupling $ \lambda(\tau) $  exhibits an interesting 
behaviour as a function of $ \tau $. 

First of all, the one-loop approximation is reliable only when $ 0
\leq \lambda(\tau) \sim \lambda_i =\lambda(\tau_i) $ or
smaller. Starting at $ \tau = 
\tau_i \; ,  \; \lambda(\tau) $ decreases until it vanishes at the first
integer multiple of $ 2\pi/\sqrt7 $. That is at $ \tau_n = 2\pi n
/\sqrt7 , \; n=0,\; 1, \; 2, \ldots $.
At these points $ h(\tau) $ becomes singular. Namely, each time
$$
\sin\left({\sqrt7\over 2}\tau \right) = 0 \quad ,\quad \tau = \tau_n =
2\pi n /\sqrt7 
$$
we have $ \lambda(\tau_n) = 0 $.
Notice that $ \lambda(\tau) $ becomes negative in the intervals
$$
\tau_n < \tau < \tau_n + \Delta_n\; ,
$$
where $ \Delta_n $ has a specific value in each interval depending 
on $\lambda_i$ and $\tau_i$ (see fig. 1). 
[$ \Delta_n $ is defined by the relation: $ h(\tau_n + \Delta_n) =
h(\tau_i) - 4\pi/\lambda_i $]. 
This shows  that the one-loop approximation ceases
to be valid at $ \tau = \tau_n + 0 $. 

The effective coupling constant $ \lambda(\tau) $ decreases with the
scale $ \tau $ in an infinite number of disjoint intervals
\be \label{97}
\tau_{n-1} + \Delta_{n-1} < \tau \leq \tau_n.
\ee

That is, we can start to run the renormalization group at $ \tau =
\tau_i $, $\tau_i$ chosen such as $\tau_{n-1} < \tau_i < \tau_{n}$, 
with a small coupling $ \lambda_i \equiv
\lambda(\tau_i) $ and keep running until $ \tau = \tau_n $. At this
point $ \lambda(\tau_n) = 0 $. In these intervals $ \tau_i,
\tau_n) $, the effective coupling  $ \lambda(\tau) $ decreases when
the space scale $ \tau $ increases as usually happens in scalar field 
theories, which is the case here (infrared stable behaviour). 

We depict in fig. 1 the running coupling constant  $ \lambda(\tau)
$ in the intervals given by eq.(\ref{97}) for $ n = 1, \; 2,  \; 3, \;
4, \; 6 $ and $ 17 $ as illustrative cases.

We see that eq.(\ref{97}) provides a {\bf hierarchy} of intervals
where the behaviour of the perturbative effective coupling [given by
eq.(\ref{lambeff})] is consistent. This hierarchy is formed by scales
following the geometric progression
\begin{eqnarray}\label{jerar}
R_0 &=& \delta \; ,\cr \cr
R_1 &=& R_0 \; e^{2\pi  /\sqrt7 }\; ,\cr \cr
\ldots && \ldots \cr \cr
R_n &=& R_0 \; e^{2\pi n /\sqrt7 } = R_0 \; [10.749087\ldots]^n\; .
\end{eqnarray}

Outside the intervals in $ \tau $ defined by eq.(\ref{97}),  the
effective coupling  $ \lambda(\tau) $ becomes negative, or very large
or both, showing that the one-loop approximation is no more reliable. 

In addition, the growth of $ \lambda(\tau) $ according to
eq.(\ref{lambeff}) when $ \tau $ decreases below $ \tau_i $
suggests that one enters here a strong coupling regime. That is, we
find an ultraviolet unstable behaviour.
We now connect this behaviour with the instability of structures and
fragmentation. 
As already established, within a given interval, the coupling $\lambda(\tau)$
grows with decreasing $\tau$, that is with a decreasing size of the system
(ultraviolet unstable behaviour). In this more and more strongly  coupled 
regime, the density fluctuations are ever larger. This means that the Jeans 
length $d_J$ decreases, the instability condition (scales larger than $d_J$)
becomes easily satisfied for smaller and smaller regions, thus  leading
to the {\bf fragmentation} of the original mass structure into substructures.
 
In the opposite regime, $\lambda(\tau)$ decreases with increasing
scale (infrared stable behaviour) and the density fluctuations decrease. 
Jeans' unstabilities thus become less probable as long we do not enter the
next interval.  

The zeroes of
$\lambda(\tau)$ determine a hierarchy of structures $ R_n = R_0 \;
e^{2\pi n /\sqrt7 } $ described above in eq.(\ref{jerar}). We thus
have a self-similar set of structures fitting one into each other.

\vskip -3cm
\centerline{\epsfig{file=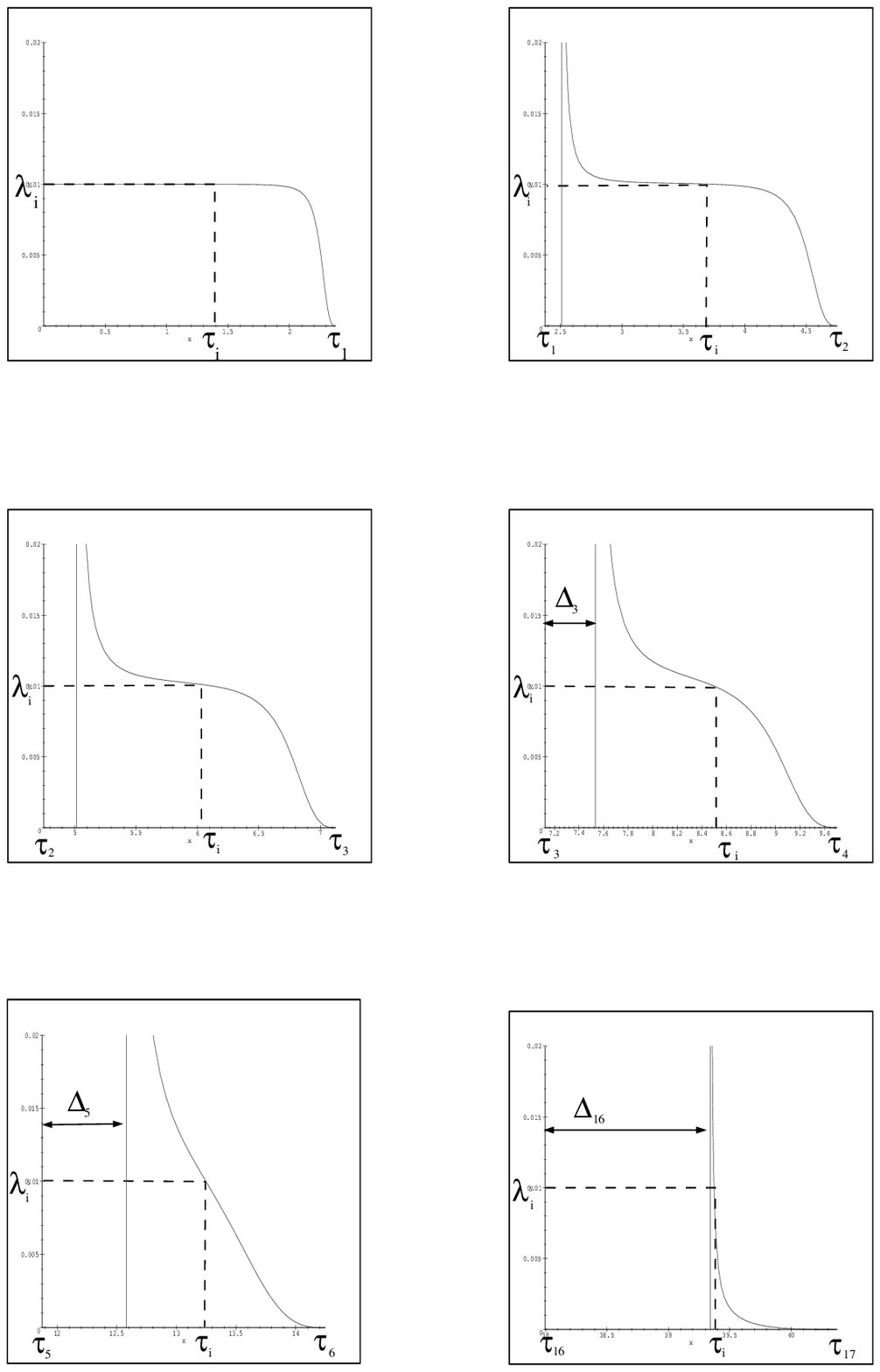,width=13.5 cm}}

\medskip
\nopagebreak[4]
\noindent
{\bf Fig 1:} {\small Above is the coupling constant $\lambda(\tau)$ picture
in 6 differents $\tau$ intervals; $[0,\tau_1]$,$[\tau_1,\tau_2]$
,$[\tau_2,\tau_3]$,$[\tau_3,\tau_4]$,$[\tau_5,\tau_6]$
and $[\tau_{16},\tau_{17}]$. The initial conditions are set such as 
$\tau_i=\tau_n-1$ and $\lambda_i=0.01$.

\newpage

\begin{appendix}
\section{Computation of the most simple Feynman diagrams}

In section 3.3.5 we have given the values of the 2-points and 3-points
averaged vertex functions. As those values are given for a one loop order
computation, only three topologocally different amputed diagrams are needed.
We will give some details here of the calculation of those diagrams.

\subsection{ Some useful integrals}

The computation of the diagrams, as we will se later, can be brought down
to that of three integrals. The calculation holds no difficulties, here are
the results:

\begin{itemize}
\item
$\displaystyle
\int_0^{-a} \!\!\!\! dx \sh(x) \sh(x+a) \,=\, {1 \over 2}
\left( a \ch(a) - \sh(a) \right)
$
\item
$\displaystyle
\int_0^{-a} \!\!\!\! dx \int_0^x \!\!\! dy \sh^2(x+a) \sh^2(y) \,=\,
{1 \over 8} \left( a^2 + a\ch(a)\sh(a) -2\sh^2(a) \right)
$
\item
$\displaystyle
\int_0^{-a} \!\!\!\! dx \sh(x+a) \sh(x) \int_0^{x+a} \!\!\!\! dy \sh^2(y)
\int_0^{x} \!\!\!dz \sh^2(z) \,=\,
$
\end{itemize}
 
\medskip
\noindent
\hfill $\displaystyle
{1 \over 16}
\left( -{a^3 \over 3} \ch(a) -{a^2 \over 2} \sh(a) - {a \over2} \sh^2(a)
\ch(a) +{4 \over 3} \sh^3(a) \right)
$

\bigskip
\noindent 
Let us see now how these integrals appear in the diagrams.

\subsection{ One loop diagrams calculation}

First is the so-called tadpole diagram, which appear in the $\phi^4$ theory
for example. We compute directly its average value.

\bigskip
\noindent
\begin{tabular}{lll}
\epsfig{file=tadpole.ps,width=1.5cm}&$=$&
$\displaystyle  C_1\int d^3 \!\!\gr{x} \int \!\! d^3\gr{x}_0\; 
\delta(\gr{x}-\gr{y})\; {\cal G}(\gr{x}-\gr{x}_0,\gr{x}-\gr{x}_0)\; 
\mu^2\;  e^{\phi_0(\gr{x}-\gr{x}_0)}$\\
&&\\
&$=$&$\displaystyle \int_{\delta}^R \!\! dr\;  \sum_l(2l+1)\;  g_l(r,r) $ \\
&&\\
&$=$&$\displaystyle
\sum_l {2l+1 \over \alpha_l \sh(\alpha_l \tau)} \int_\delta^R \!
{dr \over r}\;  \sh\left[\alpha_l \ln\left({r \over R}\right)\right]
\sh\left[\alpha_l \ln\left({r \over \delta}\right)\right] $\\
\end{tabular}
 
\bigskip
\noindent
Let us make the following change in the variables,
$x=\alpha_l\ln\left({r \over\delta}\right)$, 

\bigskip
\noindent
\begin{tabular}{lll}
\epsfig{file=tadpole.ps,width=1.5cm}&$=$&
$\displaystyle
\sum_l {2l+1 \over \alpha_l^2 \sh(\alpha_l \tau)} \int_0^{-\alpha_l
\tau}\!\!\!\!\!\!\!\! dx
\sh(x+\alpha_l\tau)\sh(x) $\\
&&\\
&$=$&$\displaystyle
\tau\sum_{l=0,l\neq 1}^{e^{\tau}} {2l+1 \over 2\alpha_l}
{\ch(\alpha_l\tau) \over \sh(\alpha_l\tau) } 
-\sum_{l=0,l\neq 1}^{e^{\tau}}{2l+1 \over \alpha_l^2}$\\
\end{tabular}

\bigskip
\noindent
The next diagram, that we call eye-like, appear in the $\phi^3$ theory. 

\bigskip
\noindent
\begin{tabular}{lll}
\epsfig{file=cercle.ps,width=2cm}&$=$&
$\displaystyle  -C_2\int d^3 \gr{x} \int  d^3\gr{x}_0\; 
{\cal G}^2(\gr{x}-\gr{x}_0,\gr{y}-\gr{x}_0)\; 
\mu^2\;  e^{\phi_0(\gr{x}-\gr{x}_0)}\; \mu^2\;  e^{\phi_0(\gr{y}-\gr{x}_0)}$\\
&&\\
&$=$&$\displaystyle -2 \int_{\delta}^R \!\!dr_1 \int_{\delta}^{R} \!\!dr_2 \sum_
l (2l+1) g_l^2(r_1,r_2) $ \\
\end{tabular}
 
\bigskip
\noindent
We split the domain of integration into two parts where $r_1$ and $r_2$ are in
a given  order, either $r_1 < r_2$ or $r_1 > r_2$. For symmetry reasons
the contribution of those parts to the result are the same. Beside we 
implement the same change in variables as in the tadpole diagram, 
$x_1=\alpha_l \ln\left({r_1 \over R}\right)$ and
$x_2=\alpha_l \ln\left({r_2 \over R}\right)$,

\bigskip
\noindent
\begin{tabular}{lll}
\epsfig{file=cercle.ps,width=2cm}&$=$&
$\displaystyle
-4 \sum_{l=0,l\neq 1}^{e^{\tau}} {2l+1 \over \alpha_l^2
\sh^2(\alpha_l\tau)}
\int_{\delta}^R {dr_1 \over r_1} \int_{\delta}^{r_1} {dr_2 \over r_2}
{1 \over \alpha_l^2 \sh^2 \alpha_l\tau } \sh^2\left(\alpha_l\ln\left(
{r_2 \over \delta}\right) \right)
\sh^2\left(\alpha_l\ln\left( {r_1 \over R}\right) \right)
$ \\
&& \\
&$=$&$\displaystyle
-4 \sum_{l=0,l\neq 1}^{e^{\tau}} {2l+1 \over \alpha_l^4
\sh^2(\alpha_l\tau)}
\int_0^{-\alpha_l\tau} \!\!\!\!\!\!\!\!\!dx_1 \int_0^{x_1} \!\!\!dx_2
\sh^2\left( x_1 +\alpha_l\tau \right)
\sh^2\left(x_2 \right)
$\\
&& \\
&$=$&$\displaystyle
- \sum_{l=0,l\neq 1}^{e^{\tau}} {2l+1 \over 2 \, \alpha_l^4
\sh^2(\alpha_l\tau)}
\left[ \alpha_l^2 \tau^2  + \alpha_l \tau \, 
\sh(\alpha_l \tau) \ch(\alpha_l \tau) \right. 
- 2\, \sh^2(\alpha_l \tau) \bigg]
$\\
&& \\
&$=$&$\displaystyle
-\frac12 \tau^2 \sum_{l=0,l\neq 1}^{e^{\tau}}
{ 2l+1 \over \alpha_l^2 \sh^2( \alpha_l \tau )} 
-\frac12 \tau \sum_{l=0,l\neq 1}^{e^{\tau}}
{ (2l+1) \ch (\alpha_l \tau ) \over
 \alpha_l^3 \sh( \alpha_l \tau )} 
+ \sum_{l=0,l\neq 1}^{e^{\tau}} { 2l+1 \over \alpha_l^4 }
 $\\
\end{tabular}
 
\bigskip

\noindent
Last, the triangle diagram, that appears in the 3-points function only,

\bigskip
\noindent
\begin{tabular}{lll}
\epsfig{file=triangle.ps,width=1.5cm}&$=$&
$\displaystyle  -C_5\int \!\!d^3 \gr{y}\int \!\!d^3 {\gr{x}} \int \!\! d^3{\gr{x}}_0
{\cal G}({\gr{x}}-{\gr{x}}_0,\gr{y}-{\gr{x}}_0)
{\cal G}({\gr{x}}-\gr{y}_0,\gr{z}-{\gr{x}}_0) $\\
&&\\
&&$\displaystyle \qquad \qquad\qquad\times \,\,
{\cal G}({\gr{x}}-\gr{z}_0,{\gr{x}}-{\gr{x}}_0)\;
\mu^2\; e^{\phi_0({\gr{x}}-{\gr{x}}_0)}\;
\mu^2\; e^{\phi_0(\gr{y}-{\gr{x}}_0)}\;\mu^2\; e^{\phi_0(\gr{z}-{\gr{x}}_0)}$ \\
&& \\
&$=$&$\displaystyle -8 \sum_l (2l+1) \int_{\delta}^R \!\!\! dr_1 \int_{\delta}^
{R} \!\!\! dr_2 \int_{\delta}^{R}\!\!\! dr_3 \; g_l(r_1,r_2) \; g_l(r_2,r_3)\;
g_l(r_3,r_1) $ \\
&&\\
\end{tabular}

\bigskip

\noindent
Here again we will split the domain into six parts where the three variable
are ordered. Contributions of theses parts are the same. And using again the
same changes in the variables:

\bigskip
\noindent
\begin{tabular}{lll}
\epsfig{file=triangle.ps,width=1.5cm}&$=$&
$\displaystyle
-48 \sum_{l=0,l\neq 1}^{e^{\tau}} {2l+1 \over \alpha_l^3
\sh^3(\alpha_l\tau)}
\int_{\delta}^R {dr_1 \over r_1} \int_{\delta}^{r_1} {dr_2 \over r_2}
\int_{\delta}^{r_2} {dr_3 \over r_3}
\sh^2\left[\alpha_l\ln\left( {r_1 \over R}\right) \right] $\\
&& \\
&&\hfill$\displaystyle
\sh\left[\alpha_l\ln\left( {r_2 \over R}\right) \right]
\sh\left[\alpha_l\ln\left( {r_2 \over \delta}\right) \right]
\sh^2\left[\alpha_l\ln\left( {r_1 \over \delta}\right) \right]
$\\
&&\\
&$=$&$\displaystyle
-48 \sum_{l=0,l\neq 1}^{e^{\tau}} {(2l+1) \over \alpha_l^6
\sh^3(\alpha_l\tau)}
\int_0^{-\alpha_l \tau} \!\!\!dx_1 \int_0^{x_1}\!\!\!dx_2
\int_0^{x_2}\!\!\! dx_3  $\\
&& \\
&&\hfill$\displaystyle
\sh^2\left(x_1+\alpha_l\tau \right)
\sh\left(x_2 \right)
\sh\left(x_2 +\alpha_l\tau \right)
\sh^2\left(x_3 \right)
$\\
\end{tabular}
 
\bigskip
 
\noindent
A trick here is to integrate $x_2$ between $0$ and $-\alpha_0\tau$
then the integrations on $x_1$ and $x_3$ disentangle,

\bigskip
\noindent
\begin{tabular}{lll}
\epsfig{file=triangle.ps,width=1.5cm}&$=$&
$\displaystyle
48 \sum_{l=0,l\neq 1}^{e^{\tau}} {2l+1 \over \alpha_l^6
\sh^3(\alpha_l\tau)}
\int_0^{-\alpha_l \tau} dx_2\;
\sh\left(x_2 \right)\;
\sh\left(x_2 +\alpha_l\tau \right) $\\
&& \\
&&\hfill$\displaystyle
\int_0^{x_2+\alpha_l \tau}\!\!\!\!\!\!dx_1
\sh^2\left(x_1\right)
\int_0^{x_2}\!\!\! dx_3
\sh^2\left(x_3 \right)
$\\
&&\\
&&\\
&$=$&$\displaystyle
3 \sum_{l=0,l\neq 1}^{e^{\tau}} {2l+1 \over \alpha_l^6
\sh^3(\alpha_l\tau)}
\left[-{\alpha_l^3 \tau^3 \over 3} \ch(\alpha_l \tau)
-{\alpha_l^2 \tau^2\over 2} \sh(\alpha_l\tau)  \right. $\\
&& \\
&&\hfill$\displaystyle
\left.
- {\alpha_l
\tau \over2} \sh^2(\alpha_l\tau)
\ch(\alpha_l \tau) +{4 \over 3} \sh^3(\alpha_l \tau) \right]
$\\
&&\\
&&\\
&$=$&$\displaystyle
-\tau^3  \sum_{l=0,l\neq 1}^{e^{\tau}}
{ (2l+1) \ch(\alpha_l \tau) \over  \alpha_l^3
 \sh^3( \alpha_l \tau )} 
- {3 \over 2}  \tau^2   \sum_{l=0,l\neq 1}^{e^{\tau}}
{ 2l+1 \over  \alpha_l^4 \sh^2( \alpha_l \tau )}
 \qquad$
\\
&&\\
&& \hfill $\displaystyle
-  {3 \over 2}\tau \sum_{l=0,l\neq 1}^{e^{\tau}}
{ (2l+1) \ch(\alpha_l \tau) \over \alpha_l^5
\sh( \alpha_l \tau )}
\,+\,4 \sum_{l=0,l\neq 1}^{e^{\tau} }
{ 2l+1 \over  \alpha_l ^6 }
$\\
\end{tabular}

It is worth to mention that the calculation of any other diagram, averaged
or not, is quite longer and require the use of formal caculus programs. 

\section{ Contribution of border terms in the renormalization group equation}

To evaluate the r.h.s of the renormalization group equation, we want to 
compute here the action on the correlation functions of the derivative:

\be {\partial \over \partial \tau}= {1 \over 2} \left(R
{\partial \over \partial R}-
\delta{\partial \over \partial \delta} \right)
\ee

According to our  Feymann rules, a diagram contributing to the n-points
vertex function, denoted $d_{x_0}(x_1,\ldots,x_n)$, with $l$ loops and $v$ 
vertex writes as follow;
 
\be
d_{x_0}(x_1,\ldots,x_n)= C ({ g \over \sqrt \mu })^{n+2(l-1)}
\int \!\!{d^3 z_1 \over z_1^2} \ldots \int \!\! {d^3 z_v \over z_v^2}
\delta(x_1-z_1) \ldots \delta(x_n-z_n) {\cal G}(z_1,.) \ldots {\cal G}(.,z_v),
\ee
where $C$ is the symetry factor of the diagram and the specific form of the 
propagators product is determined by the topological structure of the diagram.
We apply now the $\tau$ derivation and make use of our boundary conditions. We
find,

\bigskip
\noindent
$\displaystyle
{\partial \over \partial \tau} d_{x_0}(x_1,\ldots,x_n)= \sum_{\{z_i\}}
C ({ g \over \sqrt \mu })^{n+2(l-1)}$
 
\medskip
\noindent
$\displaystyle \hfill
\int \!\!{d^3 z_1 \over z_1^2} \ldots \int \!\! {d^3 z_v \over z_v^2}
\delta(x_1-z_1) \ldots \delta(x_n-z_n) {\cal G}(z_1,.) \ldots
{\partial \over \partial \tau}{\cal G}(.,.) \ldots {\cal G}(.,z_v),
$
\bigskip
where the sumation is over the derivative positions in the product. 
We then write:
\be
{\partial \over \partial \tau} {\cal G}(x,y) \equiv H(x,y)
\ee
where,
\be
H(r,r^{\prime})
= {1 \over 4 \pi \sqrt{r r^{\prime}}}
\sum_l {2l+1 \over \sh^2(\alpha_l \tau) } \left\{
\ch\left[\alpha_l \ln {r r^{\prime} \over \delta R}  \right]
\ch\left(\alpha_l \tau \right)-\ch\left[\alpha_l\ln {r^> \over
r^<} \right] \right\} P_l(\cos \gamma)
\ee
We compare now ${\cal G}$ and $H$ for $ \tau \gg 1 $ and generic
values of $ r $ and $ r^{\prime} $ when  $\tau$ is not close to its
critical values;  ${2k\pi \over \sqrt7}$
 
We can then write  ${x \over \delta}=\sqrt{ R \over \delta}(1 +
\epsilon_1)\; , {y \over \delta}= \sqrt{ R \over \delta}(1 + \epsilon_2) $.
That is, $x$ and $y$ are not close to the borders.
Then,
 
\be
g_l(x,y) = -{1 \over 4 \pi}
{1 \over \sqrt{\delta R}} {2l+1 \over 2 \alpha_l}
\left[1+(\alpha_l-1)\epsilon_1- 
(1+\alpha_l)\epsilon_2+ O(\epsilon_1^2,\epsilon_2^2,\epsilon_1
\epsilon_2) \right]
\ee
\be
h_l(x,y) \sim {1 \over  \pi} {2l+1 \over \sqrt{\delta R}} \,
e^{-\tau \alpha_l}\left[ 1+\alpha_l\epsilon_1+\alpha_l\epsilon_2+
O(\epsilon_1^2,\epsilon_2^2,\epsilon_1 \epsilon_2)\right]
\ee
In this second equation $h_l$ is defined by analogy with $g_l$.
So, in this region,
\be
g_l(x,y) \gg h_l(x,y)\quad l \neq 0
\ee
and
\be
g_0(x,y) \sim h_0(x,y)
\ee
Then,
\be
{\cal G}(x,y) \gg H(x,y)
\ee
This means that, in this region (i.e not close to the borders), and for
non critical values of $\tau$, the border terms are subdominant.
 
\end{appendix}

\end{document}